\documentclass[twocolumn,showpacs,nofootinbib]{revtex4}

\usepackage{epsf}
\usepackage{graphicx} 
\usepackage{amssymb}
\usepackage{dcolumn}
\usepackage{amsmath}
\usepackage{hyperref}

\def\apj{ApJ\,}                 
\def\apjl{ApJL\,}                
\def\apjs{ApJS\,}               
\def\mnras{MNRAS\,}             
\def\aap{A\&A\,}                
\def\nat{Nature\,}              
\def\araa{ARA\&A\,}             

\def\jcap{JCAP\,}

\def\thetaB{\mbox{\boldmath$\theta$}}
\def\thetazB{\mbox{\boldmath$\theta_0$}}

\def\lsim{~\rlap{$<$}{\lower 1.0ex\hbox{$\sim$}}}
\def\gsim{~\rlap{$>$}{\lower 1.0ex\hbox{$\sim$}}}

\begin{document}

\title{The Impact of Magnification and Size Bias on Weak Lensing Power Spectrum \\ and Peak Statistics}

\author{Jia Liu$^{1}$}
\email {jia@astro.columbia.edu}

\author{Zolt\'an Haiman$^{1,2}$}
\email {zoltan@astro.columbia.edu}

\author{Lam Hui$^{2,3}$}
\email {lhui@astro.columbia.edu}

\author{Jan M. Kratochvil$^{4,5}$}
\email {jank@physics.miami.edu}

\author{Morgan May$^{6}$}
\email {may@bnl.gov}

\affiliation{ {$^1$ Department of Astronomy and Astrophysics, Columbia University, New York, NY 10027, USA}} 
\affiliation{ {$^2$ Institute for Strings, Cosmology, and Astroparticle Physics (ISCAP), Columbia University, New York, NY 10027, USA}} 
\affiliation{ {$^3$ Department of Physics, Columbia University, New York, NY 10027, USA}} 
\affiliation{ {$^4$ Department of Physics, University of Miami, Coral Gables, FL 33146, USA}}
\affiliation{ {$^5$ Astrophysics and Cosmology Research Unit, University of KwaZulu-Natal, Westville, Durban, 4000, South Africa}   }
\affiliation{ {$^6$ Physics Department, Brookhaven National Laboratory, Upton, NY 11973, USA}   }

\date{\today}

\begin{abstract}
  The weak lensing power spectrum is a powerful tool to probe
  cosmological parameters. Additionally, lensing peak counts contain
  cosmological information beyond the power spectrum.  Both of these
  statistics can be affected by the preferential selection of source
  galaxies in patches of the sky with high magnification, as well as
  by the dilution in the source galaxy surface density in such
  regions.  If not accounted for, these biases introduce systematic
  errors for cosmological measurements.  Here we quantify these systematic
  errors, using convergence maps from a suite of ray-tracing N-body
  simulations.  At the cut-off magnitude $m$ of on-going and planned major weak lensing surveys,
  the logarithmic slope of the cumulative number counts 
  $s\equiv d\log n(>m)/d\log m$ is in the range $0.1\lsim s \lsim0.5$.
  At $s \approx 0.2$, expected in the $I$ band for LSST, the inferred values of
  $\Omega_m$, $w$ and $\sigma_8$ are biased by many $\sigma$ (where
  $\sigma$ denotes the marginalized error) and therefore the
  biases will need to be carefully modeled.  We also find that the
  parameters are biased differently in the ($\Omega_m$,
  $w$, $\sigma_8$) parameter space when the power spectrum and when
  the peak counts are used. In particular, 
  $w$ derived from the power spectrum is less affected than $w$ derived from
  peak counts, while the opposite is
  true for the best-constrained combination of
  $\sigma_8\Omega_m^\gamma$ (with $\gamma=0.62$ from the power
  spectrum and $\gamma = 0.48$ from peak counts).  
  This suggests that the combination of the power spectrum and peak counts 
  can help mitigate the impact of magnification and size biases.
\end{abstract}


\maketitle

\section{Introduction}\label{Introduction}

By measuring the distortions of background galaxy shapes by
foreground masses (galaxies, galaxy clusters, and large-scale
structures), weak gravitational lensing (WL) surveys probe the mass
density fluctuations throughout the cosmic span (see recent reviews by
\cite{Refregier2003,Schneider2005,Hoekstra2008,Bartelmann2010}). WL
observations, in conjunction with cosmological simulations, can be
used to place precise constraints on cosmological parameters. Recent
WL surveys, such as COSMOS \cite{Schrabback2010} and
CFHTLenS \cite{Kilbinger2013}, have measured the shear power spectrum
and have already placed useful constraints on $\Omega_m$ (the
matter density of the universe), $\sigma_8$ (the amplitude of the
primordial power spectrum on a scale of 8$h^{-1}$ comoving Mpc), and
$w$ (the dark energy equation of state).

Because of the statistical nature of WL surveys, it is important to
have an unbiased sample of source galaxies, fairly sampling the
foreground density fluctuations across the sky.  In this paper, we
investigate possible sources of bias in flux-limited surveys, arising
from a preferential selection of source galaxies in patches of the sky
with high magnification, as well as by the dilution in the source
galaxy surface density in such regions (known as magnification bias;
hereafter MB).  MB has been studied extensively in the past for its
impact on galaxy--quasar and galaxy--galaxy correlation functions in
2D \cite{Webster1988, Fugmann1988, Narayan1989, Schneider1989,
  Villumsen1995, Villumsen1997, Moessner1998, Kaiser1998,
  Loverde2007}, and in 3D \cite{Matsubara2000,Hui2007,Hui2008}, and on
the statistics of the Lyman-$\alpha$ forest \cite{Loverde2010}.  An
additional size bias (hereafter SB) can be present in surveys in which
the selection of the source galaxies depends on their angular sizes.
If not accounted for, these biases represent a systematic error for
cosmological measurements. In the context of WL, the impact of MB and
SB have been studied for the power spectrum
\cite{Schmidt2009,Schmidt2009b} and for high peaks caused by
individual NFW halos \cite{Schmidt2011}.

Ref.~\cite{Schmidt2009} has shown that ignoring MB and SB in the
shear power spectrum can cause $2-3\sigma$ deviations in cosmological
parameter estimation for a DETF~\cite{DETF} Stage III experiment, such
as the Dark Energy Survey.\footnote{\url{http://www.darkenergysurvey.org}} Future
WL surveys with larger sky coverage and/or deeper observations, such
as those by planned by the Large Synoptic Survey Telescope\footnote{\url{http://www.lsst.org}}
(LSST) and
Euclid\footnote{\url{http://sci.esa.int/euclid}}, will have significantly
better statistical sensitivity, and therefore can be more severely
impacted by these biases.

In this paper, we first show that MB is indeed significant for the
power spectrum, extending earlier results~\cite{Schmidt2009} to
explicitly compute the biases on cosmological parameters.  We then
focus on the impact of MB on peak counts. Lensing peaks were first
considered as a cosmological probe in early ray-tracing simulations a
decade ago \cite{JV00}.  Peak counts have received increasing
attention in recent years \cite{Dietrich2010, Maturi2010,
Kratochvil2010, Yang2011, Marian2012, Kratochvil2012, Pires2012,
Yang2013, Bard2013} as a way to access cosmological information from
the strong non-Gaussianities in the lensing fields.  In particular,
these studies have shown that the number and height-distribution of
peaks have high cosmological sensitivity, and can improve cosmological
constraints by a factor of $\sim$two, compared to using the power
spectrum alone.

Peak counts are a simple and robust statistic, defined by recording
local maxima in a 2D shear or convergence ($\kappa$) map,
smoothed by suitable filters. Ref.~\cite{Yang2011} investigated the
physical origin of the individual $\kappa$ peaks, by tracing their
contributing light rays back in time across their N-body simulation
boxes. They found that high peaks (with amplitudes $\gtrsim$
3.5$\sigma_{\kappa}$, where $\sigma_{\kappa}$ is the r.m.s. of the
convergence~$\kappa$) are typically created by individual massive
halos.  It has been shown that MB increases the signal-to-noise and
therefore the total number of such peaks~\cite{Schmidt2011}.  By
comparison, low peaks ($\sim1-2\sigma_{\kappa}$) are typically caused
by a combination of (cosmology-independent) shape noise and a
(cosmology-dependent) constellation of 4--8 lower-mass halos.  These
halos have masses of a few $\times10^{12}~{\rm M_\odot}$, and are
offset by $\sim$arcmin from the line of sight to the center of the
peak.  The low peaks are especially promising, as they carry the
majority of cosmological information, and are relatively insensitive
to baryonic cooling that affect the halo cores \cite{Yang2013}.  We
therefore extend the earlier results of \cite{Schmidt2011} 
for high peaks,  where increases in both the peak heights and
number of high peaks were seen,  to the low peaks, and to 
explicitly compute the  biases on cosmological parameters.

To study the impact of MB, we build a simple numerical model to derive
cosmological parameters (and their error bars) using either the power
spectrum or peak counts measured in our simulations. We then apply
magnification bias to a set of ``true'' convergence maps (which
faithfully represent the projected dark matter distribution in a
fiducial flat $\Lambda$CDM cosmology) to create mock ``biased'' maps,
mimicking an observed dataset. For each of these ``biased''
datasets, we find the best-fit set of the three cosmological
parameters ($\Omega_m, w$ and $\sigma_8$), using the ``true'' maps for
the model fitting.  Finally, we quantify the difference between the
inferred cosmology and the true fiducial cosmology, as a function of
the strength and sign of the magnification and size bias (determined
by the slope of the galaxy luminosity function and the galaxy size
distribution).

The rest of this paper is organized as follows: in \S~\ref{MB}, we
introduce the formalism of magnification bias, and discuss its
principal ingredient, the galaxy luminosity function.  We then
describe our computation methods in \S~\ref{Methodology}, including
the convergence maps created with our ray-tracing N-body simulations,
computing the power spectra and the peak distributions from these
maps, determining the cosmology-dependence of these quantities, and
finally applying biases to the maps to create mock observations. We
present our main results in \S~\ref{Results}, where we fit the mock
data, and show that MB and SB will indeed alter the derived
cosmological parameters by many $\sigma$. Finally, in
\S~\ref{Summary}, we summarize our conclusions and the implications of
this work.

\section{Magnification Bias}\label{MB}

Gravitational lensing causes a bias by modulating the apparent surface
density of galaxies on the sky, through two competing
effects~\cite{Turner1984}. First, lensing can magnify (or demagnify)
individual source galaxies in the background, increasing (or
decreasing) their total flux.  In a flux-limited WL surveys, 
some otherwise excluded faint galaxies can
therefore make it into (or drop out of) the sample because of this
(de)magnification. Second, a similar (de)magnification applies to the
patch of the sky around the galaxy, geometrically diluting (or
enhancing) the apparent surface density of galaxies in this region.
These two effects counteract each other, and the net bias depends on
the slope of the intrinsic (unlensed) galaxy luminosity function at
the survey flux limit.  In addition to these effects, lensing can
increase (or decrease) the apparent angular size of spatially resolved
individual galaxies.  If either the survey selection, or a derived
statistic such as WL shear, depends on the apparent size, then this
can introduce an additional size bias.

\subsection{Formalism}

To quantify the effect of MB, we follow the discussion in Appendix A
of \cite{Hui2007}.  Including the effect of lensing on both the flux
and on the geometrical surface density, we have the relation
\begin{eqnarray}\label{nzdef2}
n(\thetaB) = n_g (\thetaB) \left[ 1 + (5s - 2) \kappa(\thetaB) \right],
\end{eqnarray}
where $n(\thetaB)$ is the observed (lensed) galaxy number density at
position $\thetaB$, as viewed by the observer, $n_g (\thetaB)$ is the
intrinsic (unlensed) galaxy number density, $s$ is the slope of the cumulative
number counts evaluated at $m_{\rm lim}$, and $\kappa(\thetaB)$ is
the convergence. This equation assumes the weak lensing limit
($\kappa\ll 1$), neglects the correspondingly small difference
$\delta\thetaB$ between lensed and unlensed directions on the sky, and
also assumes that galaxy number density fluctuations $\delta n_g/n_g$
on the angular scales of interest are small, as well.  Under these
assumptions, the above equation is valid to first order in
$\kappa,\delta\thetaB$, and $\delta n_g$.  Finally, if we assume a survey
 with a sharp magnitude cut-off at $m_{\rm lim}$:

\begin{eqnarray}
\label{sdef2}
s = \left.\frac{\partial {\rm log}_{10} n_g} {\partial m} \right|_{m_{\rm lim}}.
\end{eqnarray}

\subsection{Galaxy Luminosity Functions and WL Surveys}

\begin{figure}
\includegraphics[scale=0.6]{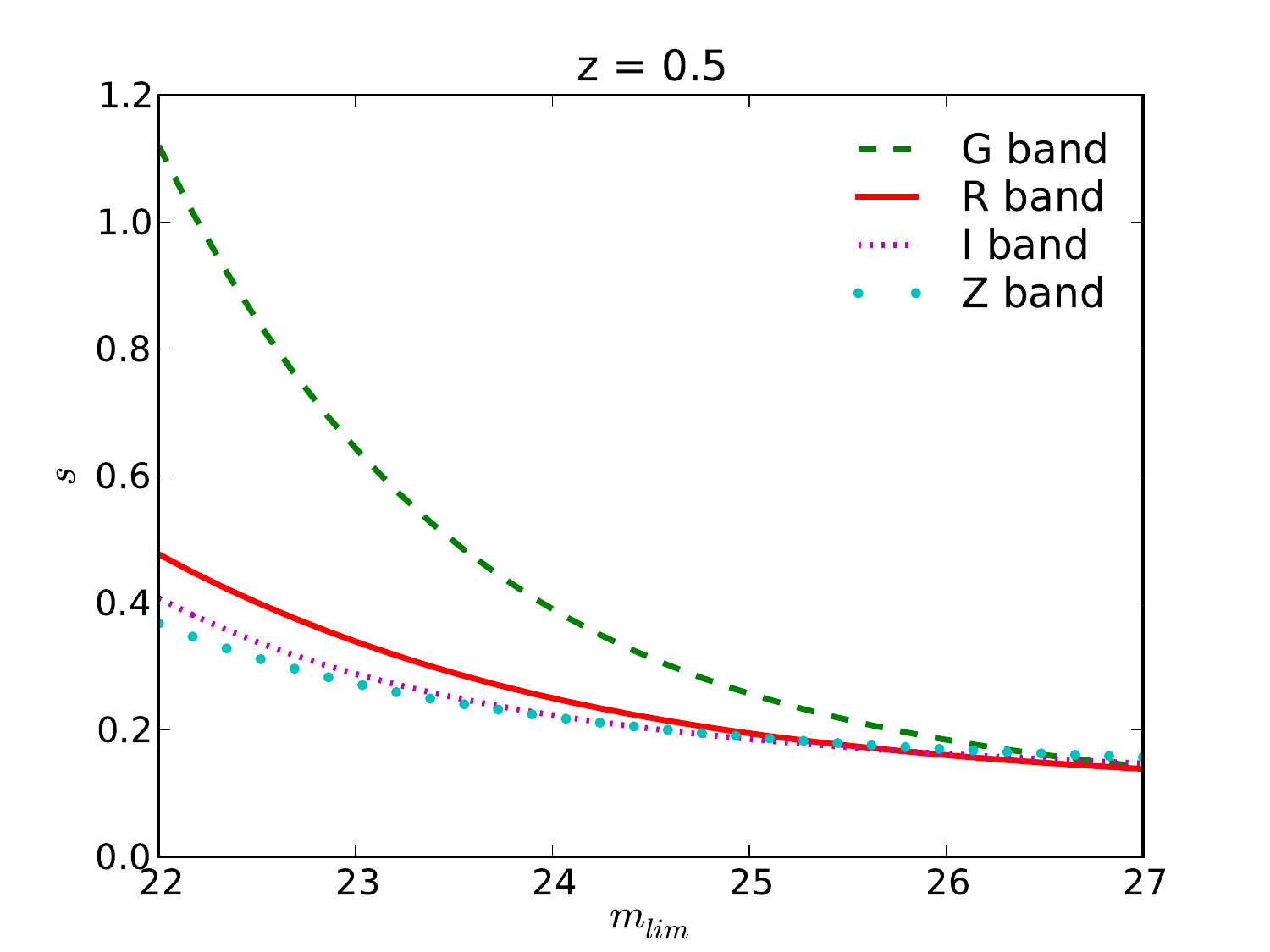}
\includegraphics[scale=0.6]{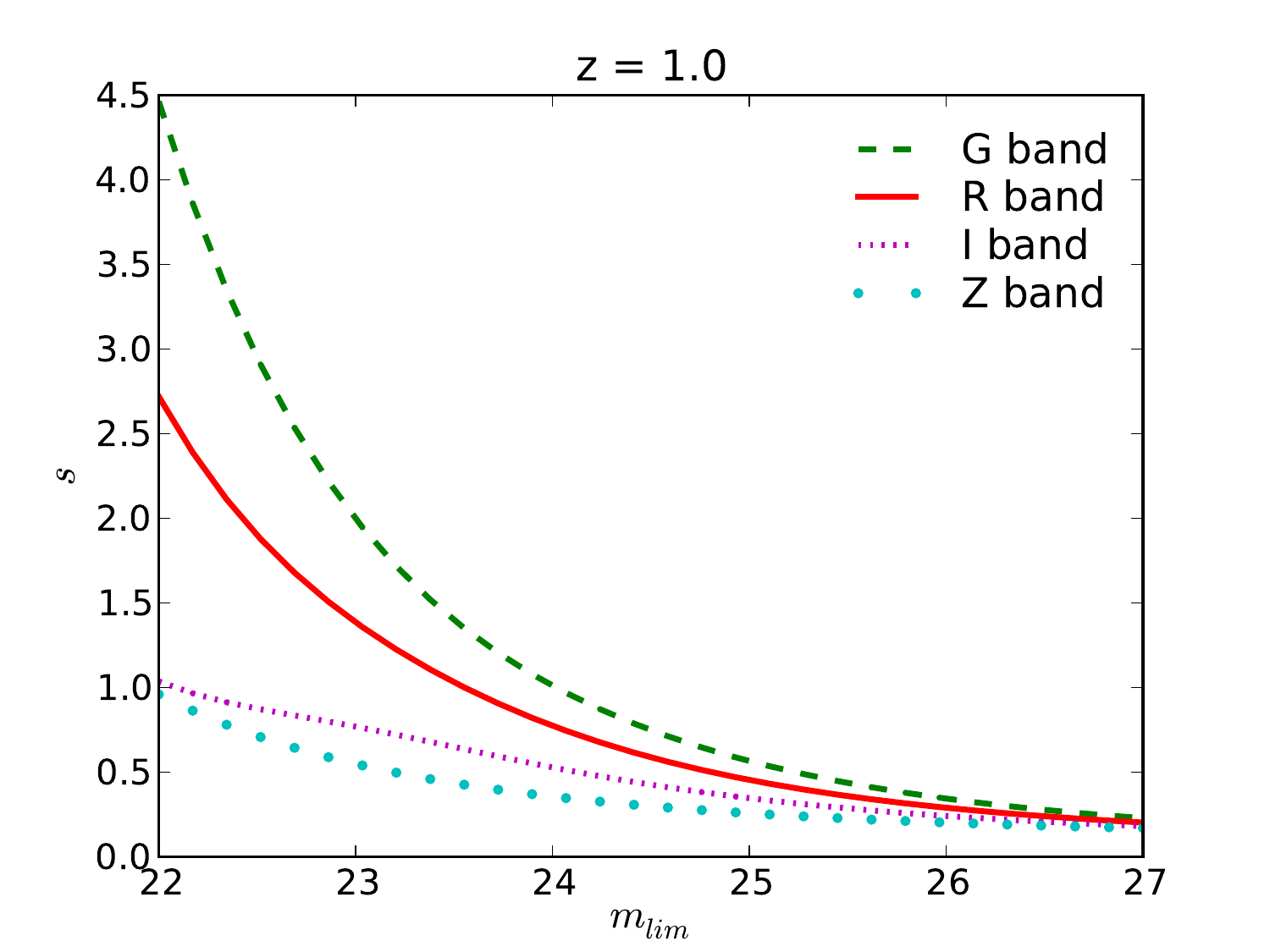}
\caption{\label{s_vs_Mlim} The logarithmic slope $s$ of the galaxy
number counts as a function of cut-off magnitude $m_{\rm lim}$ at two
different redshifts $z=0.5$ (top) and $z=1$ (bottom). Four different
filters (G, R, I, Z) are shown. WL surveys target a depth of $m_{\rm
lim} > 24$ to achieve a sufficiently large galaxy number density. As a
result, the relevant range of $s$ is $0.1\lsim s \lsim 0.6$ (see
Table~\ref{tab:relevant s}).}
\end{figure}

The magnitude of MB depends on the galaxy luminosity function through
the logarithmic slope $s$.  Observed luminosity functions are well
described by a Schechter function \cite{Schechter1976},
\begin{eqnarray}
\label{LF eqn}
\Phi(M) dM &=& (0.4 \ln10) \Phi^{\star} \left[ 10^{-0.4(M-M^{\star})} \right]^{\alpha+1} \\ 
&\times&\exp \left[ -10^{-0.4(M-M^{\star})}\right] dM \nonumber,
\end{eqnarray}
where $\Phi(M) dM$ is the number density of galaxies with magnitude
between $M$ and $M+dM$, $\Phi^{\star}$ is a characteristic number
density (in Mpc$^{-3}$), and $M^{\star}$ is a characteristic
magnitude.  It consists of a power-law with index $\alpha$ at the
faint end and a exponential cut-off at the bright end. The cumulative
galaxy number density can be written as
\begin{eqnarray}
\label{n_g}
n_g=\int_{-\infty}^{M_{lim}} \Phi(M) dM.
\end{eqnarray}
Note that this equation holds at a given redshift. In
Fig.~\ref{s_vs_Mlim}, we show $s$ calculated using
equation~(\ref{sdef2}), as a function of cut-off magnitudes in the G,
R, I and Z bands at redshifts $z=0.5$ and $z=1$. In this calculation,
we used the measurements of $\Phi^{\star}, M^{\star}$ and $\alpha$ by
\cite{Gabasch2004, Gabasch2006}, which are all redshift-dependent.
\footnote{Eq.1 in ref.~\cite{Gabasch2004} describes the redshift evolution 
of $\Phi^{\star}, M^{\star}$ and $\alpha$. The parameters can be found in 
table 3 \& 4 of  ref.~\cite{Gabasch2004} for the G band, and in table 9 
(``case~3'', with a constant $\alpha=-1.33$)  of ref.~\cite{Gabasch2006} 
for the R, I, Z bands.}

Table~\ref{tab:relevant s} lists the magnitude limits and the
corresponding values of $s$ for several current and future WL
surveys. We note that in order to measure the shape of the galaxies,
it is necessary to adopt a brighter magnitude than for the point
sources. For surveys where the magnitude limit was available only for
point sources, we adopted a one magnitude brighter value for
$m_{\rm lim}$.  For simplicity, for broad multi-band filters (R+I+Z),
we have calculated $s$ using the central I band.  Table~\ref{tab:relevant
s} shows that surveys with $m_{lim,I} \approx 24-25$ have $s\approx0.2$,
assuming a mean redshift $z=0.5$. While the effect of MB almost 
disappears ($s\approx0.4$) for galaxies at $z=1.0$. Ref.~\cite{Schmidt2009}
has shown that the $z$--dependence is much weaker than the $s$--dependence.
For LSST, we expect the effective galaxy number density after applying lensing 
cuts to peak at a lower redshift ($z=0.5-0.8$) than the raw sample ($z\gsim1.0$). 
To illustrate the effect of MB, we adopt $s=0.2$ as our fiducial value, 
corresponding to $z=0.5$ for the conservative cut (see Figure 7 in ref.~\cite{Chang2013}).
 We will show in \S~\ref{Results} that MB
 will significantly affect the power spectrum and the peak counts at this value.

\begin{table}
\begin{tabular}{lcccc} 
\hline					
	&	Magnitude limit	&	$s (z=0.5)$	&	$s (z=1.0)$	&	ref	\\
\hline									
LSST	&	I $\le$ 24.8	&	0.19	&	0.38	&	\cite{LSSTSciBook2009}	\\
Euclid	&	R+I+Z $\le$ 24.5	&	0.20	&	0.43	&	\cite{Euclid2009}	\\
COSMOS	&	I $\le$ 25	&	0.19	&	0.35	&	\cite{COSMOS2007}	\\
CFHTLS	&	I $\le$ 24.7	&	0.19	&	0.39	&	\cite{CFHTLS2012}	\\
DES	&	I $\le$ 24.3	&	0.21	&	0.46	&	\cite{DES2005}	\\
DUNE	&	R+I+Z $\le$ 24.5	&	0.20	&	0.43	&	\cite{DUNE2009}	\\
KiDS	&	R $\le$ 25.2 (*)	&	0.24	&	0.69	&	\cite{KiDS2013}	\\
HSC	&	I $\le$ 26.2 (*)	&	0.18	&	0.32	&	\cite{HSC2010}	\\
\hline
\end{tabular}
\caption[]{\label{tab:relevant s} Magnitude limits and corresponding
$s$ (number count slope) at $z=0.5$ and 1.0 for current and future WL surveys.
 For surveys in which only the point source magnitude limit was
available (marked by a ``*''), we reduced $m_{\rm lim}$ by 1 magnitude
to represent an extended source magnitude cut. In the broad multi-band
(R+I+Z), we calculated $s$ for the central I band.}
\end{table}

\subsection{Size Bias}

If a survey has a cut in galaxy size $r$, in addition to a flux
cut, then equation~(\ref{nzdef2}) is modified to
\begin{eqnarray}\label{nzdef2size}
n(\thetaB) = n_g (\thetaB) \left[ 1 + (5s + \beta - 2) \kappa(\thetaB) \right],
\end{eqnarray}
where in the case of a sharp cut, the new term $\beta$ is the logarithmic slope of the galaxy size
distribution,
\begin{eqnarray}
\label{bdef2}
\beta = -\left.\frac{\partial {\rm ln} n_g} {\partial {\rm ln} r} \right|_{r_{\rm lim}}.
\end{eqnarray}
This equation assumes that the size and flux cuts are independent, and
also that the slopes $s$ and $\beta$ only weakly depend on $r$ and
$M$.  Under these simple assumptions, the effects of size and
magnification bias are equivalent, and only the combination $(5s +
\beta)$ matters.  A more sophisticated treatment will eventually be
necessary (and will depend on the details of the survey, including how
galaxy sizes affect measurement errors). Here we simply note that at
the limiting magnitudes of $m_{lim}\approx 24-25$, the observed angular size
distribution has a slope of $\beta\sim 3$
\cite{Schmidt2009b}. Therefore the additional effect of size bias is
equivalent to increasing the value of $s$ by $0.6$, i.e. the relevant
fiducial value for LSST with a flux + size cut is changed from
$s\approx 0.2$ to $s\approx 0.8$. This means that the sign of the
 effect changes when we add size bias to magnification bias,
 as the effect of galaxy density dilution dominates over individual
 galaxy magnification.

\section{Methodology}\label{Methodology}

\subsubsection{N-body Simulations}

The N-body simulations and lensing maps were created with the
Inspector Gadget lensing simulation pipeline on the New York Blue IBM
BlueGene supercomputer. The N-body simulations are the same as the
ones used in our earlier work \cite{Kratochvil2010, Yang2011,
  Kratochvil2012, Yang2013, Bard2013}. We refer readers to these
papers for more detailed information. Here we briefly describe the
basis of the simulations and the parameters used.

This work uses in total 35 different N-body simulations, covering 7
different cosmological models (1 fiducial cosmology plus 6
variations), each with 5 independent realizations of the same input
primordial power spectrum. We chose our fiducial cosmological model to
be $\Omega_m=0.26$, $w=-1.0$, Hubble constant $H_0=0.72$, with a
primordial matter power spectrum with $\sigma_8=0.798$ and a spectral
index of $n_s=0.96$, using the best fit values from the seven-year
results by the {\it WMAP} satellite \cite{Komatsu2011}. We vary each
of the 3 parameters ($\Omega_m, w$ and $\sigma_8$) one at a time (a
higher value and a lower value than in the fiducial model), while
keeping the other 2 parameters at the fiducial values. The 6
non-fiducial models have values of $\Omega_m = \{0.23, 0.29\}$ (while
$\Omega_{\Lambda}=\{0.77, 0.71\}$ to keep a spatially flat universe),
$w = \{-1.2, -0.8\}$ and $\sigma_8 = \{0.75, 0.85\}$. The combinations
are listed in Table~\ref{tab:Cosmologies}.

\begin{table}
\begin{tabular}{l|c|c|c} 
\hline
 & $\sigma_8$ & $w$ & $\Omega_m$  \\
\hline
Fiducial &  0.798 &  -1.0 & 0.26 \\
High-$\sigma_8$ & 0.850 & -1.0 & 0.26\\
Low-$\sigma_8$ & 0.750 & -1.0 & 0.26\\
High-$w$ & 0.798 & -0.8 & 0.26\\
Low-$w$ & 0.798 & -1.2 & 0.26\\
High-$\Omega_m$ & 0.798 & -1.0 & 0.29\\
Low-$\Omega_m$ & 0.798 & -1.0 & 0.23\\
\hline
\end{tabular}
\caption[]{\label{tab:Cosmologies} Cosmological parameters in each model. 
  The universe is assumed to be spatially flat ($\Omega_\Lambda+\Omega_m=1$).}
\end{table}

The N-body simulations were generated using a modified version of the
Gadget-2 code\footnote{\url{http://www.mpa-garching.mpg.de/gadget/}},
and they consist of dark matter only. Each run has a box size of $240
h^{-1}$ comoving Mpc, containing $512^3$ particles. This corresponds
to a mass resolution of $7.4\times10^9h^{-1}M_\odot$. The initial
(linear) total matter power spectrum was computed with the
Einstein-Boltzmann code CAMB\footnote{\url{http://camb.info/}}
\cite{Lewis2000} at $z=0$ and scaled back to $z=100$, which is the
starting point of our simulations. The power spectrum was then fed
into N-GenIC, the initial condition generator associated with Gadget2.

\subsubsection{Ray-Tracing and Lensing Maps}

To construct convergence maps, we perform ray-tracing. First, we
output 3D boxes at redshifts corresponding to every $\sim$ 80 Mpc
(comoving). We then divide the 3D box into many parallel pieces and
project each slice onto a 2D plane perpendicular to the observer's
line of sight, using the the triangular shaped cloud (TSC) scheme
\cite{Hockney1988}. In the next step, we convert the surface density
to gravitational potential at each plane using Poisson's
equation. Each 2D plane has a resolution of $4096\times4096$ pixels.
We then follow $2048\times2048$ light rays from $z=0$, traveling
backward through the projection planes. The deflection angle and WL
convergence and shear are calculated at each plane for each light
ray. These depend on the first and second derivatives of the
gravitational potential, respectively. Between the planes, the light
rays travel in straight lines. Finally, for each of the 7 cosmological
models, we create 1,000 convergence maps of 12 deg$^2$ each in
size. This is done by mixing simulations of different realizations,
and randomly rotating and shifting the simulation data cubes.

We add galaxy ellipticity noise to our maps, due to variations in the
intrinsic shapes of galaxies, and their random orientations on the
sky. This shape noise is added to the raw convergence maps using a
redshift-dependent expression for the noise in one component of the
shear \cite{Song2004}:
\begin{equation}
\sigma_\lambda(z)=0.15+0.035z.
\end{equation}
For each pixel, we add $\kappa_{noise}$ drawn from a random Gaussian
distribution centered at zero with variance \cite{vanWaerbeke2000}
\begin{equation}
\label{eq:galaxynoise}
\sigma^2_{\rm noise}=\frac{\langle\sigma^2_\lambda\rangle}{n_{\rm gal} \Delta\Omega},
\end{equation}
where $n_{\rm gal}$ is the number of galaxies per ${\rm arcmin^{2}}$, and
$\Delta\Omega$ is the solid angle of a pixel in units of ${\rm
  arcmin^2}$. In the case of LSST, we expect $n_{gal} \sim 30$
arcmin$^2$\cite{Chang2013} for galaxies that are usable for shape measurements,
and it follows that $\sigma_{\rm noise}=0.33$. This is much larger
than the WL signal, whose r.m.s. value (at $z=1$) for noise-free maps
is $\sigma_{\kappa} = 0.02$. To average out the random galaxy noise,
we perform smoothing on individual maps with a Gaussian kernel:
\begin{eqnarray}
\label{kappa_Gdef}
\kappa_G(\thetazB)&=&\int\,d^2\theta W_G(|\thetaB-\thetazB|)\kappa(\thetaB)
\\
W_G(\theta)&=&\frac{1}{2\pi\theta^2_G}\exp(-\frac{\theta^2}{2\theta^2_G})\label{eq-WG},
\end{eqnarray}
where $\kappa_G$ is the smoothed $\kappa$ value at pixel $\theta_0$
and $W_G$ is the Gaussian kernel with a smoothing scale
$\theta_G=1/\sqrt 2$ arcmin.\footnote {We note that in previous
  papers of this series \cite{Kratochvil2010, Yang2011,
    Kratochvil2012, Yang2013}, a different definition of
  $W_G(\phi)=\frac{1}{\pi\theta^2_G}\exp(-\frac{\phi^2}{\theta^2_G})$
  was adopted.  When using the more commonly used definition $W_G$
  (eq.~\ref{eq-WG}), our smoothing scale of $\theta_G=1/\sqrt 2$
  arcmin is equivalent to their $\theta_G=1$ arcmin.}  The choice of
smoothing scale has a known effect on the total peak counts and the
shape of the peak distribution. Increasing the smoothing scale
generally reduces the total number of peaks and increases the width of
the distribution. It has been shown that smaller smoothing scales
($\sim~1/\sqrt 2$ arcmin) generally give better constraints, and also
that combining a few different scales can further improve the errors
\cite{Kratochvil2012, Marian2012}.
Finding the range of optimal smoothing scales and filter shapes will
have to be done specifically for each survey with different
characteristics. In this paper, we continue to use the single
smoothing scale $\theta_G=1/\sqrt 2$ arcmin for simplicity, and to
facilitate comparison with previous works.

For simplicity, we use only convergence maps for source galaxies at
the single redshift $z=1$, as the $z$--dependence of MB
has shown to be weak \cite{Schmidt2009}.
Future work should employ tomography with multiple
redshifts, and fold into the analysis the actual $z$--distribution of
the source galaxies.  In total, we have 7,000 convergence maps; we
call these the ``true'' maps, since they do not include any
magnification bias. We use this set of maps to predict the
cosmology-dependent observables (power spectra or peak counts), which
will be described in detail in \S~\ref{CosmoModel}.

\subsubsection{Power Spectra and Peak Counts}

The power spectrum is the most widely used statistic in current WL
surveys, and has already been shown to be affected significantly by
MB~\cite{Schmidt2009}. We revisit the impact of MB on the power
spectrum in order to cross-check our simulation results, and to
explicitly compute the resulting biases on the cosmological
parameters.

We first compute the power spectra for 
spherical harmonic index $\ell$ in the range $100<\ell<100,000$, with
1000 equally spaced (linear) bins.
This covers the range of angles from our pixel size ($\sim 6$ arcsec)
to the linear size of our maps ($\sim 3.5$ deg).  In our previous work
\cite{Kratochvil2010,Kratochvil2012} we compared our numerical power
spectrum with the semi-analytical power spectrum obtained using the
Limber approximation \cite{Limber1953} and integrating the nonlinear
3D matter power spectrum along the line of sight \cite{Smith2003}. Our
power spectrum loses power on large scales below $\ell \sim 400$ due
to our finite box size, and on small scales above $\ell\sim20,000$ due
to spatial resolution; there is excellent agreement with the
semi-analytic predictions between these scales.

Peak counting is done by simply scanning through the pixels on a
convergence map, and identifying local maxima (i.e. pixels with a
higher value of $\kappa$ than its surrounding 8 pixels). We then
record the number of peaks as a function of their central $\kappa$
value.

\subsubsection{Applying Bias to the Convergence Maps}

On each of the 1000 maps in our fiducial cosmology, we apply different
levels of MB, ranging from $s=-0.5$ to 1.0, with a step size $\Delta
s=0.01$. To do this, on each fiducial map, we take into account the
$(5s-2)\kappa$ factor in eq.~(\ref{nzdef2}) and add $\kappa_{noise}$ 
when smoothing the
map. Eq.~(\ref{kappa_Gdef}) becomes (with $\thetaB$ dependence
suppressed for $\kappa$ and $\kappa_{noise}$):
\begin{eqnarray}
  \kappa_G=\frac {\int\,d^2\theta W_G \left[\left(1+\left(5s-2\right)\kappa\right)
      \kappa+\kappa_{noise}\right]} {\int\,d^2\theta W_G \left[1+\left(5s-2\right)\kappa\right]}.
\end{eqnarray}

This is the smoothed $\kappa$ at each pixel, weighted by the galaxy
number densities modified by MB. Note that we assume the intrinsic
(unlensed) galaxy number density to be a constant - this ignores the
effects of shot noise arising from a discrete sampling of the $\kappa$
field by a finite number of galaxies, as well as the clustering of
galaxies.  Other than applying MB, the same procedures are then
followed to add noise, smooth the maps, count peaks, or compute power
spectra, on the ``bias'' maps, as for the ``true'' maps.

\subsubsection{Predictions in Other Cosmologies}\label{CosmoModel}

In this subsection, we describe how we interpolate (and extrapolate)
the peak counts and power spectra for other cosmologies, using our set
of simulations in the 7 different cosmologies listed in
Table~\ref{tab:Cosmologies}.

First, for individual convergence maps, we histogram the $\kappa$
peaks into 200 equally spaced bins ranging from $\kappa=-0.02$ to
0.19 (this choice for the number of bins will be justified in \S~\ref{Discussions}(vii) below). 
We then calculate the mean peak distribution (average of the
1,000 maps) in each of the 7 cosmology models. To predict the peak
distribution for an arbitrary combination of cosmological parameters,
we treat each $\kappa$ bin individually, and use a Taylor expansion:
\begin{eqnarray}
\label{eq:Derivatives}
\overline{N_i} (\Omega_m, w, \sigma_8)&=&
\overline{N_i}(\Omega_m^\star, w^\star, \sigma_8^\star) \\ \nonumber
&+&\frac{\partial \overline{N_i}}{\partial \Omega_m} \Delta \Omega_m
+\frac{\partial \overline{N_i}}{\partial w}\Delta w
+\frac{\partial \overline{N_i}}{\partial \sigma_8}\Delta \sigma_8.
\end{eqnarray}
Here $\overline{N_i}$ denotes the total number of peaks in the $i^{\rm
  th}$ bin ($i$=1,~2...200), averaged over 1000 maps. $\Delta\Omega_m,
\Delta w$ and $\Delta\sigma_8$ are the differences of the desired
parameters ($\Omega_m, w, \sigma_8$) from the fiducial parameters
($\Omega_m^*, w^*, \sigma_8^*$). 

The same method was followed for the power spectrum, by simply
replacing the peak counts $N_i$ with $P_i=P(\ell_i)$, the total power in
the $i^{th}$ $\ell$ bin.

In the body of our paper below, we chose to use the fiducial and the
``high'' models as defined in Table~\ref{tab:Cosmologies}, to compute
the cosmology derivatives in equation~(\ref{eq:Derivatives}) by a
simple finite difference.  We call these ``forward derivatives''.
Given that we also have ``low'' models for each parameter, ideally we
could use all three models to refine these predictions, either by
including second-order terms in the Taylor expansion, or using
two-sided linear derivatives.  In practice, we chose to avoid a
second--order expansion, in order to be able to perform an analytical
$\chi^2$ minimization (see next subsection).  We have attempted to use
a two-sided derivative, but have found that this caused numerical
problems (the discontinuity in the derivative can cause the fitting
procedure, described below, to become stuck).  We therefore use the
forward derivatives in the bulk of this paper. We will discuss the
differences in our results if we use ``backward derivatives'' instead
in \S~\ref{Discussions}.

\subsubsection{Finding the Best-Fit Cosmology}\label{CosmoModelFits}

To fit a cosmology to one of our biased maps (or more generally to an
arbitrary peak count distribution), we minimize a $\chi^2$, defined as
\begin{equation}\label{chisqeq}
\chi^2(\Omega_m, w, \sigma_8)=\Delta N_i C_{ij}^{-1} \Delta N_j.
\end{equation}
Here $\Delta N_i=N_i^{\prime}-\overline{N_i}(\Omega_m, w, \sigma_8)$
is the difference between the peak distribution in a given single map
($N_i^{\prime}$) and the model ($\overline{N_i}$) in the $i^{th}$ bin,
and $C_{ij}^{-1}$ is the unbiased estimator of the
inverse covariance matrix \cite{Hartlap2007, Anderson2003}. Summation
is implied over repeated indices $i$, $j$. We make the simple
assumption that the peak counts depend linearly on the three
parameters. It then becomes possible to write down an analytical
solution to the best fitted parameters. By defining
\begin{eqnarray}
X_{i\alpha}&=&\frac{\partial \overline{N_i}}{\partial p_{\alpha}}  \\
Y_i&=&N_i^{\prime}-\overline{N_i}(\Omega_m^\star, w^\star, \sigma_8^\star),
\end{eqnarray}
where $p_{\alpha}=(\Omega_m, w, \sigma_8$) is a three-component vector
and $\alpha=1,2,3$ denotes one of the three parameters, we can rewrite
\begin{eqnarray}
\Delta N_i&=&Y_i-X_{i\alpha}dp_{\alpha}\\
\chi^2&=&\left(Y_i-X_{i\alpha}dp_{\alpha}\right) C_{ij}^{-1} \left(Y_j-X_{j\beta}dp_{\beta}\right)
\end{eqnarray}

Setting $d\chi^2/d(dp_\alpha)=0$, we obtain
\begin{eqnarray}
&&X_{i\alpha}C_{ij}^{-1}\left(Y_j-X_{j\beta}dp_{\beta}\right)+\nonumber \\
&&\left(Y_i-X_{i\beta}dp_{\beta}\right) C_{ij}^{-1} X_{j\alpha} =0
\end{eqnarray}
which is symmetric in $i$ and $j$, and hence the two terms can be
written combined as
\begin{eqnarray}
X_{i\alpha}C_{ij}^{-1}\left(Y_j-X_{j\beta}dp_{\beta}\right)=0
\end{eqnarray}
and the difference between the best fit and the fiducial model is
simply
\begin{eqnarray}
\label{analytical dp}
dp_{\beta}=(X_{i\alpha}C_{ij}^{-1}X_{j\beta})^{-1} (X_{i\alpha} C_{ij}^{-1} Y_j).
\end{eqnarray}

To check these analytical calculations and to eliminate potential
numerical errors from matrix inversion, we also directly minimized
eq.~(\ref{chisqeq}) using the numerical {\it scipy} routine
``optimize.minimize''.\footnote{\url{http://scipy.org/}} These
numerically identified best-fits were nearly indistinguishable from
the analytical calculations above. For convenience and to keep
computational costs to a minimum, we used the analytic approach in our
main calculations.

The same fitting procedure was performed using the power spectrum, by
simply replacing the peak count $N_i$ with the power spectrum
$P_i=P(\ell_i)$ in the above equations. In the case of the power
spectrum model (as for the peaks), we used the covariance matrix
derived using noisy maps, to include the higher power at small $\ell$
induced by the galaxy shape noise. However, to measure the power 
spectrum derivatives with respect to cosmological parameters, we 
computed $dP$ using the noiseless maps directly (since noise adds 
linearly). We choose to use the noisy maps directly, but only with 
$100<\ell<20,000$, as cutting off at $\ell=20,000$ (corresponding to $\sim1$ arcmin) 
is equivalent to smoothing but has the advantage of faster computation.

The above procedure, applied to each of the 1,000 individual ``bias''
maps, returns a set of 1,000 best-fit parameters for each specific
value of $s$.  We then use the distribution of these best-fits to find
the average bias in the cosmology parameters (corresponding to the
mean best fit), confidence levels, and the goodness-of-fit values.

\section{Results}\label{Results}

\subsection{Power Spectrum}\label{Power Spectrum}

\begin{figure} 
\includegraphics[scale=0.45]{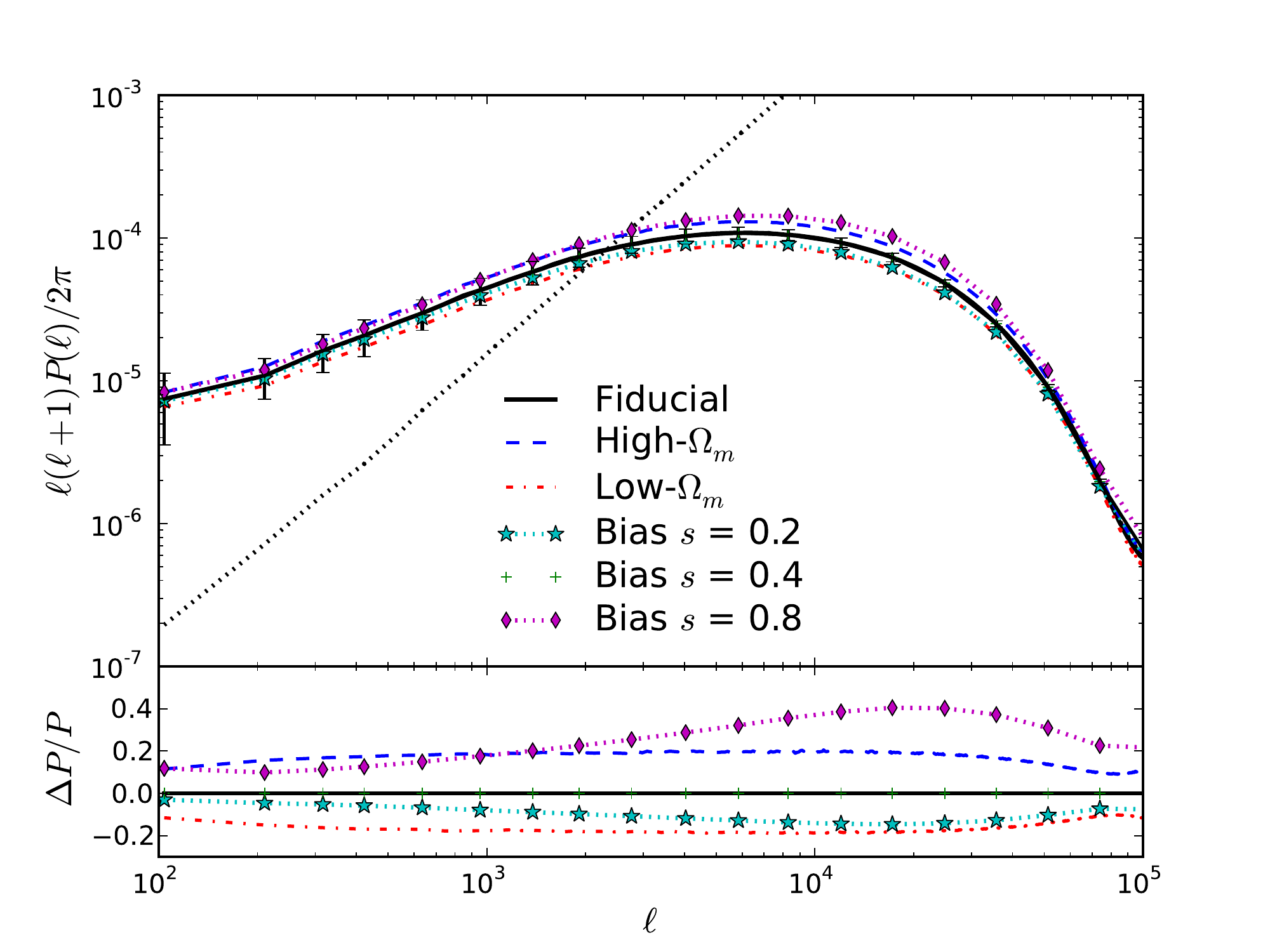}
\includegraphics[scale=0.45]{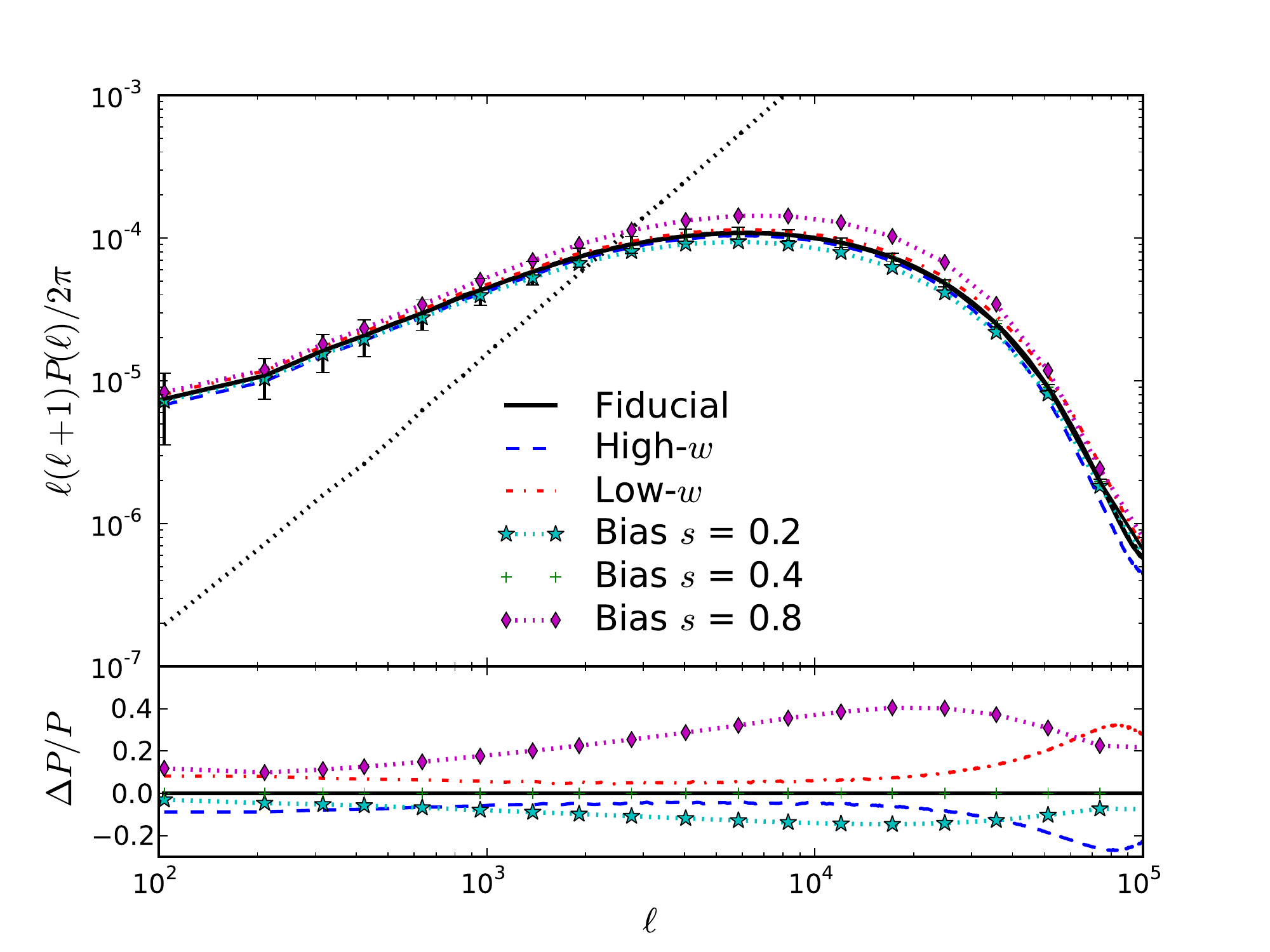}
\includegraphics[scale=0.45]{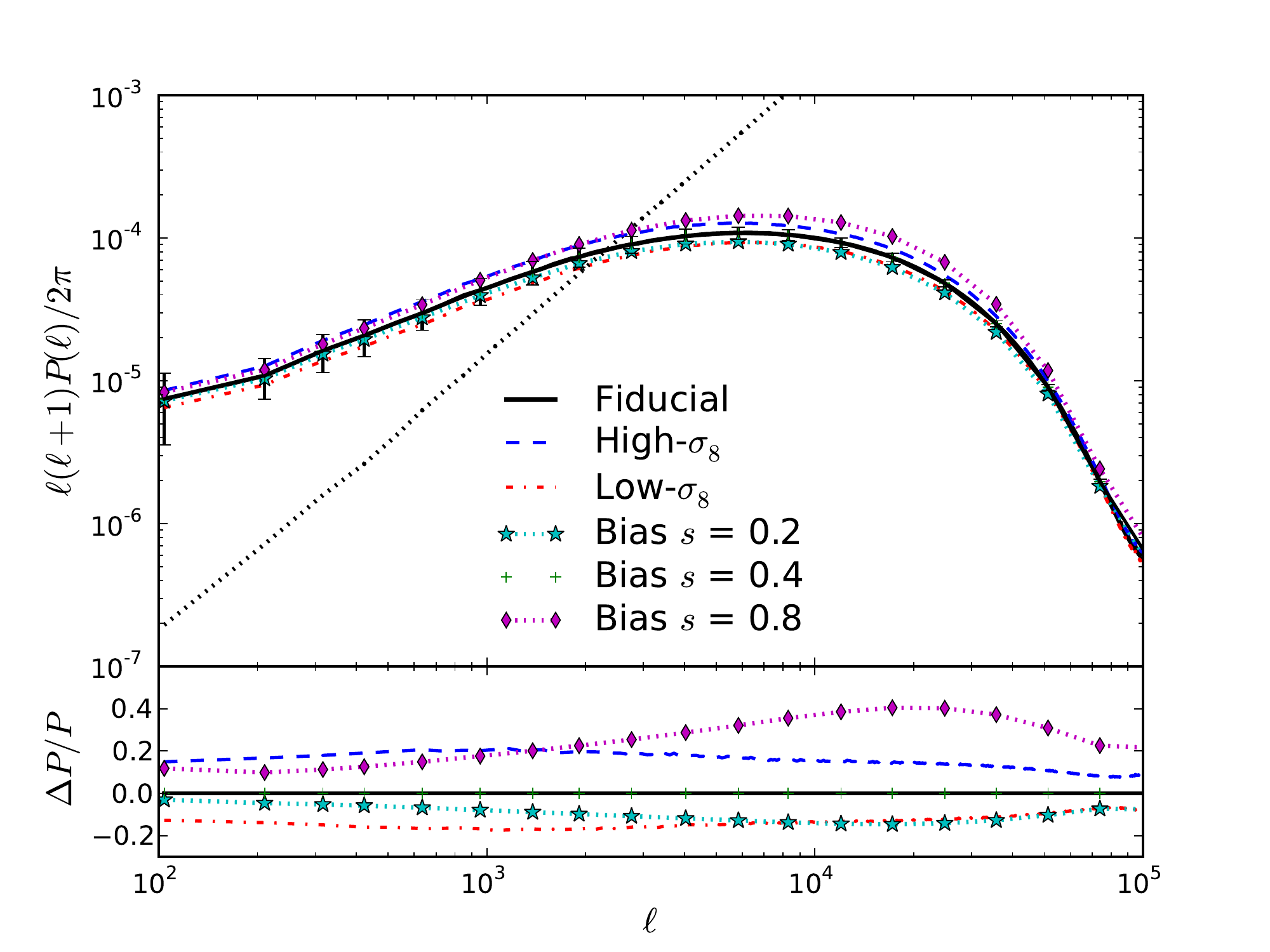}

\caption{\label{powspec} Changes in the convergence power spectrum
  caused by magnification bias, as well as by varying individual
  cosmological parameters. Three levels of bias on the fiducial model are shown with $s=0.2,
  0.4$ and 0.8. From top to bottom, besides the fiducial model, we
  also show changes due to variations in $\Omega_m$(top), $w$ (middle)
  and $\sigma_8$ (bottom). Error bars are for a 12 deg$^2$ sky; we
  expect them to decrease by a factor of $\sim 40$ after scaling the
  results to LSST's 20,000 deg$^2$ survey. The black dotted line is the 
  galaxy noise for $n_{gal}=30$ arcmin$^2$. }
\end{figure}

The impact of MB on the power spectrum is illustrated in
Fig.~\ref{powspec}. The levels of bias we chose to show are $s=0.2$,
0.4 and 0.8. The value $s=0.2$ is close to that expected in LSST; 
$s=0.4$ is the special case when MB effect
disappears completely ($q\equiv5s-2=0$); and $s=0.8$ corresponds to
$q=1$ in \cite{Schmidt2009}, close to the value expected in the
presence of an additional size bias. For comparison, we also show the
impact on $P(\ell)$ of varying each cosmological parameter.

\begin{figure} 
\includegraphics[scale=0.45]{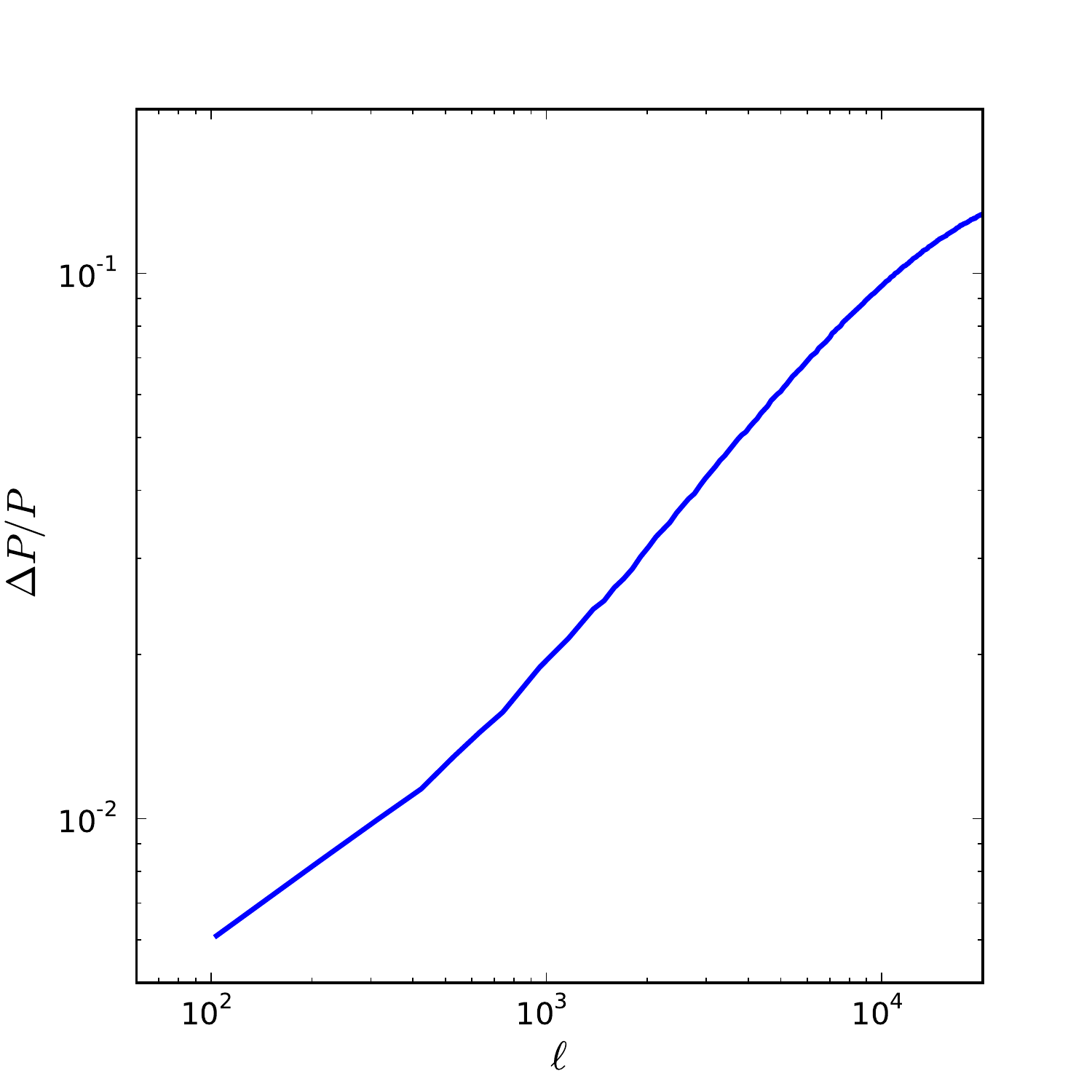}
\caption{\label{shearpowspec} Fractional difference of shear power
 due to bias ($s=0.8$). The slope and values of this curve are very close
 to the ones obtained by ref.~\cite{Schmidt2009} (their Fig.~1)
 in the range $1,000<\ell<10,000$.
 Our $s=0.8$ case is equivalent to their $q=1$ case, as they 
 also included the reduced shear correction.}
\end{figure}

For $s\approx 0.2$, the observations suffer a negative bias magnitude
of $q=-1.0$. In this case, the effect of diluting a patch of sky wins
over the number density increase due to magnification. At all $\ell$
bins, the power is reduced, as the result of the decreasing $\kappa$
fluctuations. For $s=0.4$, we have $q=0$ and expect the MB effect to
be absent. This is verified by the lack of any difference between the
power spectrum in the $s=0.4$ and the fiducial (unbiased) models, and
merely serves as a test of our numerical code. For $s=0.8$, the power
is increased on all scales; this behavior has the opposite sign of the
$s=0.2$ case, and is consistent with the expectations from
$q=5s-2=2>0$. For cross-check, we calculate $\Delta P / P$ for $s=0.8$ using 
shear maps. Our results (Fig.~\ref{shearpowspec}) are very close to
 the ones obtained by ref.~\cite{Schmidt2009} (their Fig.~1) in the range $1,000<\ell<10,000$
(note that our $s=0.8$ case is equivalent to their $q=1$ case, as they 
also included the reduced shear correction). However, we noticed
that the amplitude of $\Delta P / P$ is 10 times smaller than if 
we use convergence maps (as in this work).

\subsection{Peak Counts}\label{Peakcounts}

\begin{figure} 
\includegraphics[scale=0.45]{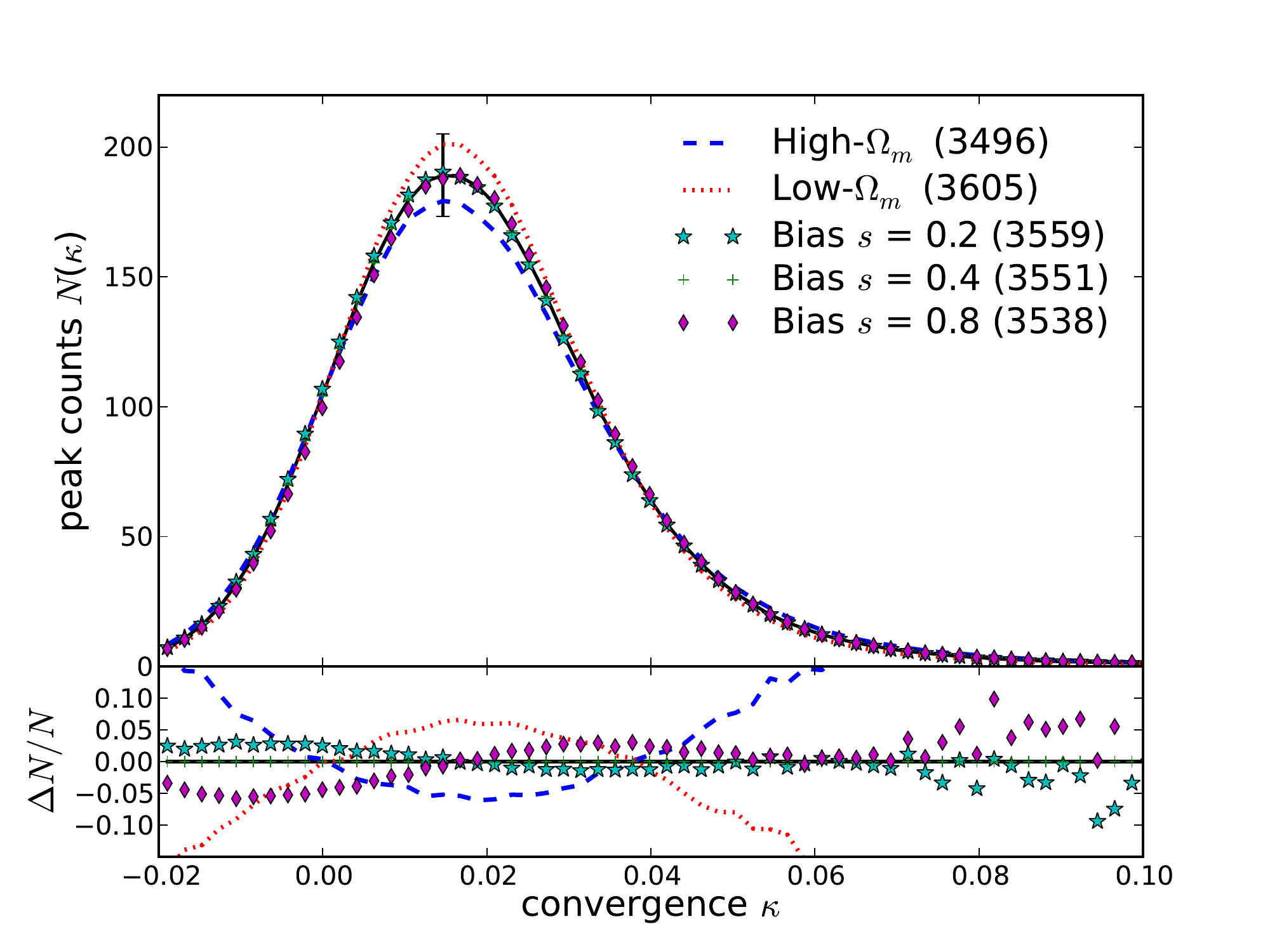}
\includegraphics[scale=0.45]{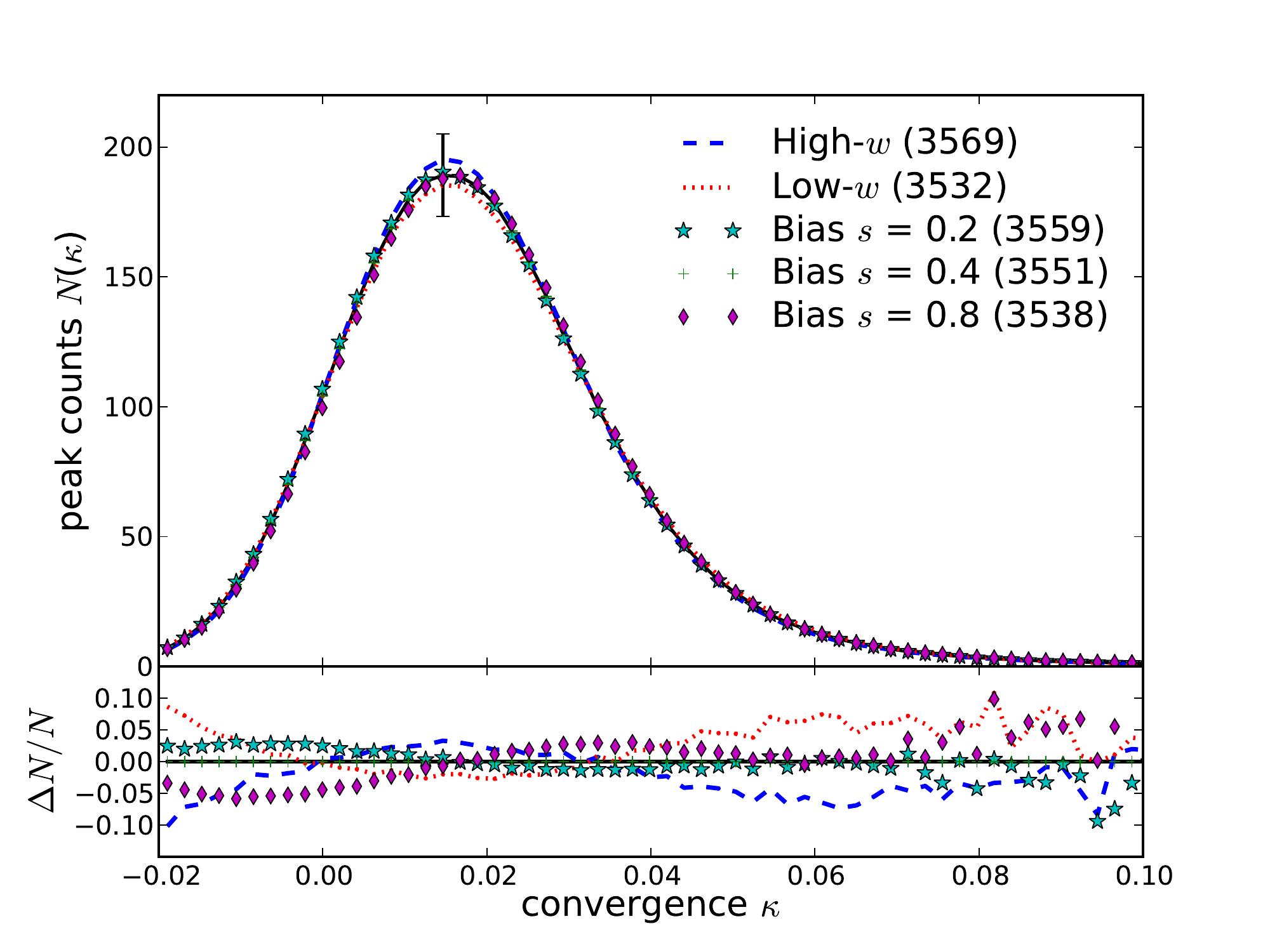}
\includegraphics[scale=0.45]{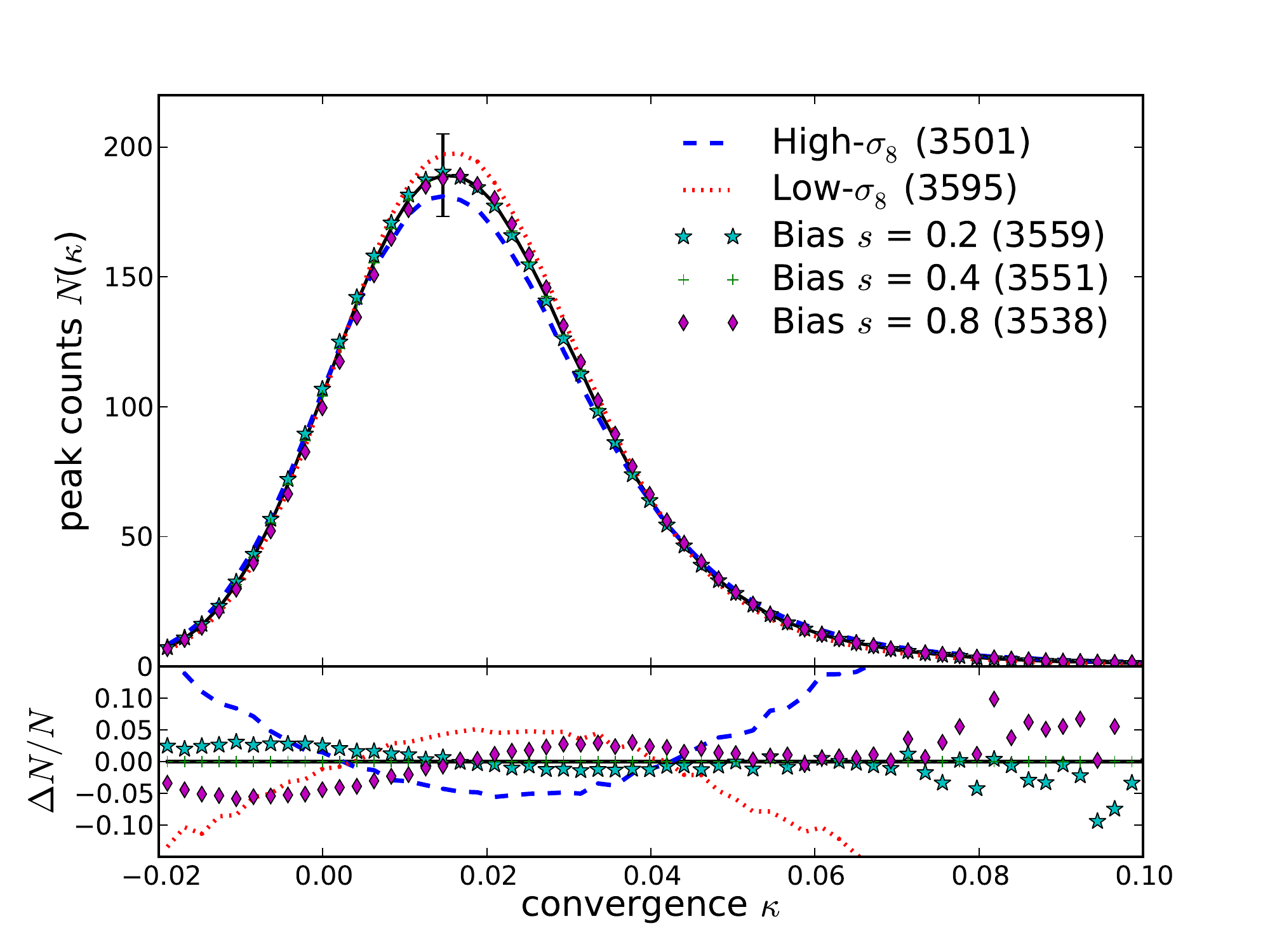}
\caption{\label{peakcounts} Peak count changes due to varying levels
  of magnification bias, as well as due to varying cosmological
  parameters. Three levels of bias on the fiducial model are shown with $s=0.2, 0.4$ and
  0.8. As in Fig.~\ref{powspec}, we also show changes due to
  variations in $\Omega_m$, $w$, and $\sigma_8$. The number in
  brackets is the total number of peaks. One error bar is shown to
  represent a typical error size for 12 deg$^2$ sky; we expect this to
  decrease by a factor of $\sim 40$ after scaling to an LSST-like
  survey.}
\end{figure}

Fig.~\ref{peakcounts} shows the impact of MB on peak counts. 
For the pure MB case of $s=0.2$, the height of any positive $\kappa$
peak is reduced due to the negative overall bias. The case $s=0.4$
continues to show no effect from MB.
Finally, for the MB + SB case of $s=0.8$
(or $q=5s-2=2$), all $\kappa$ peaks are boosted to a higher value,
and consequently the whole distribution is shifted toward the right.
The peak counts change in a direction opposite to the $s=0.2$ case,
and with a larger amplitude, as expected.  We note that for this large
positive bias, the abundance of the $\gsim 3\sigma$ (or $\kappa\gsim
0.06$) peaks increases (as discussed in ref.~\cite{Schmidt2011}), but
the number of the low peaks is {\em reduced}.

\begin{figure} 
\includegraphics[scale=0.35]{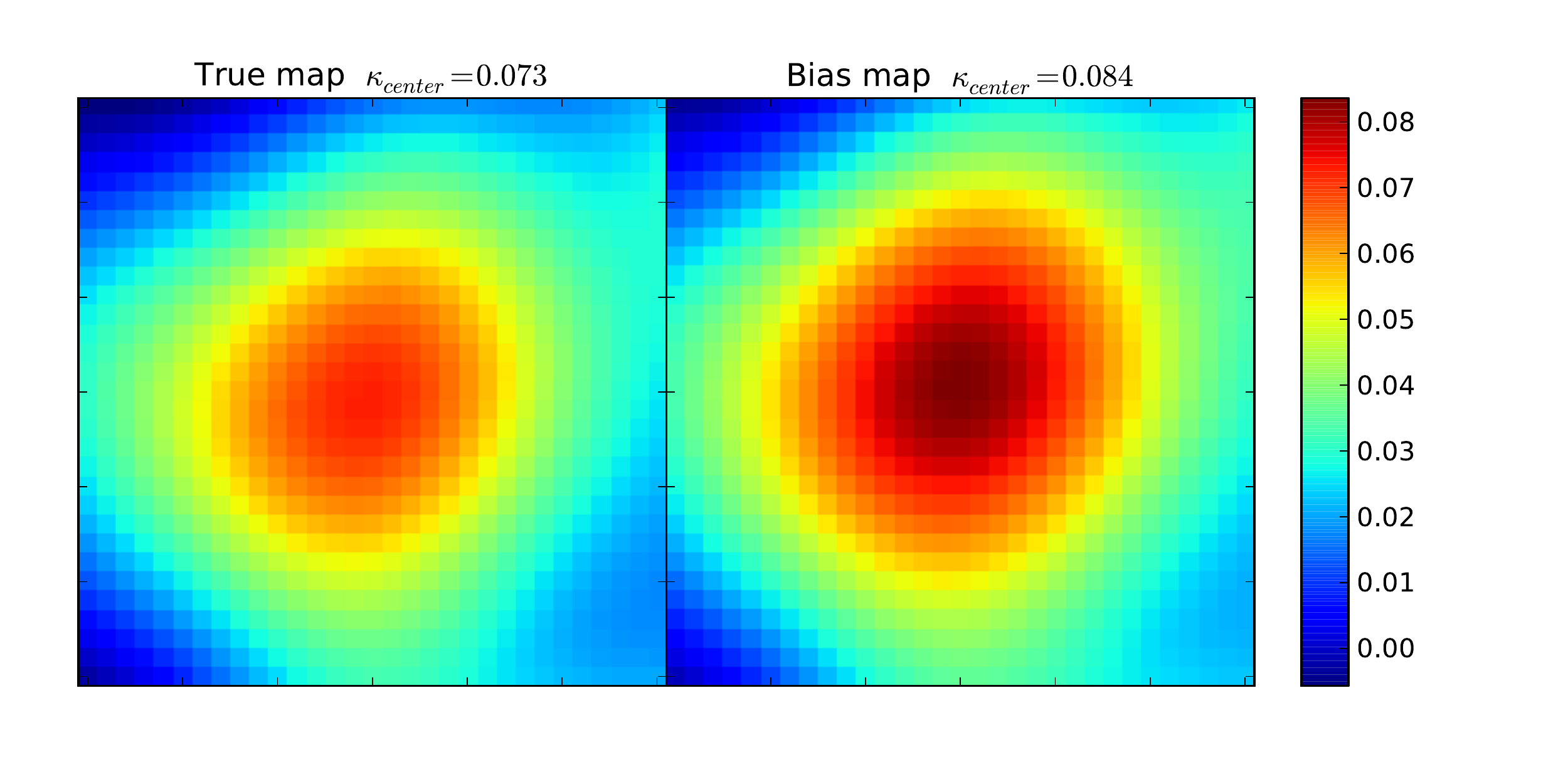}
\caption{\label{samplehipeak} An example of a high peak (central pixel
  of the true map at left panel) for $s=1.5$. Most high peaks are
  characterized by their relatively round shape, due to one single
  massive halo. After a positive magnification bias is applied to the
  map (right panel), the peak remains, and with a higher $\kappa$
  value.}
\end{figure}

\begin{figure} 
\includegraphics[scale=0.35]{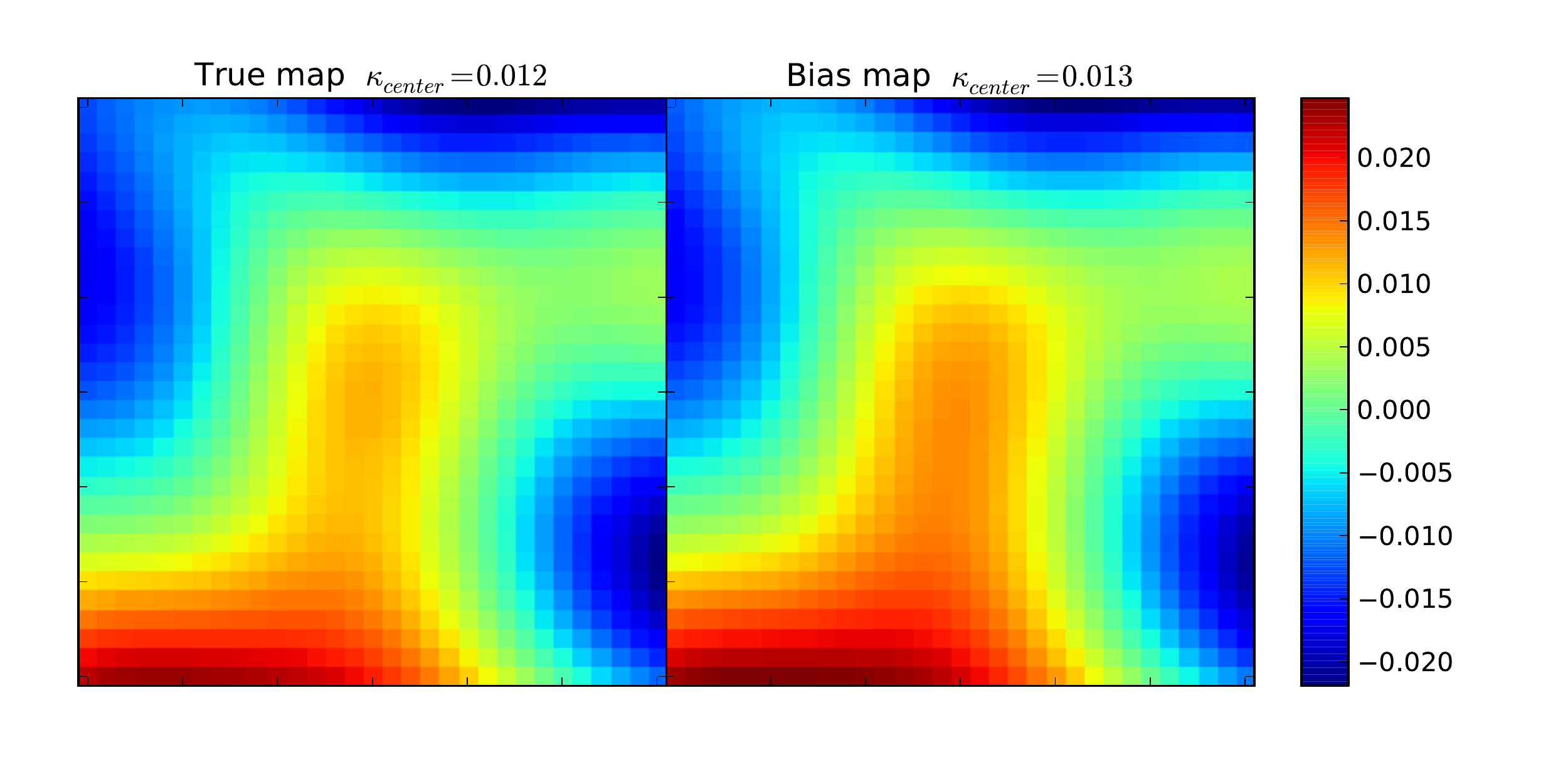}
\caption{\label{sample disappear peaks} An example of a typical low
  peak (central pixel of the true map at left panel) that disappears
  when a positive magnification bias is applied with $s=1.5$. Such
  peaks are normally found to be adjacent to another peak with a
  somewhat higher height, or between multiple higher peaks. After
  magnification bias (right panel), this particular low peak merges,
  through a ``ridge'', with its neighboring peak.}
\end{figure}

A positive MB effect ($s>0.4$) also reduces the total number of peaks
(the number in brackets in Fig.~\ref{peakcounts}). By directly
comparing an example of the ``bias'' maps against its original
``true'' version, we found that out of the $\sim3600$ peaks in total,
$\sim 120$ peaks disappeared after MB, while only $\sim 60$ new peaks
were created. By visual examination of the maps, we found that peak
disappearance and creation tends to happen in complex regions, where
many peaks are interconnected through filament-like structures.  As an
illustration of this, in Fig.~\ref{samplehipeak} we show a typical
``high'' peak. The shape of this peak is fairly round, likely due to
one single massive halo. High peaks like this normally remain a peak
after MB. In contrast, Fig.~\ref{sample disappear peaks} shows a
typical low peak that disappears after MB is applied. The original low
peak merges into the neighboring, somewhat higher-amplitude peak at
the lower left corner -- this can be attributed to the lensing bias
creating a ``ridge'' between the two original peaks.
The opposite phenomenon happens when the overall MB is negative
($s<0.4$), where we see an increase in total number of peaks, due to
the bias ``destroying'' ridges and causing a net increase in the
number of low peaks.

\begin{figure} 
\includegraphics[scale=0.45]{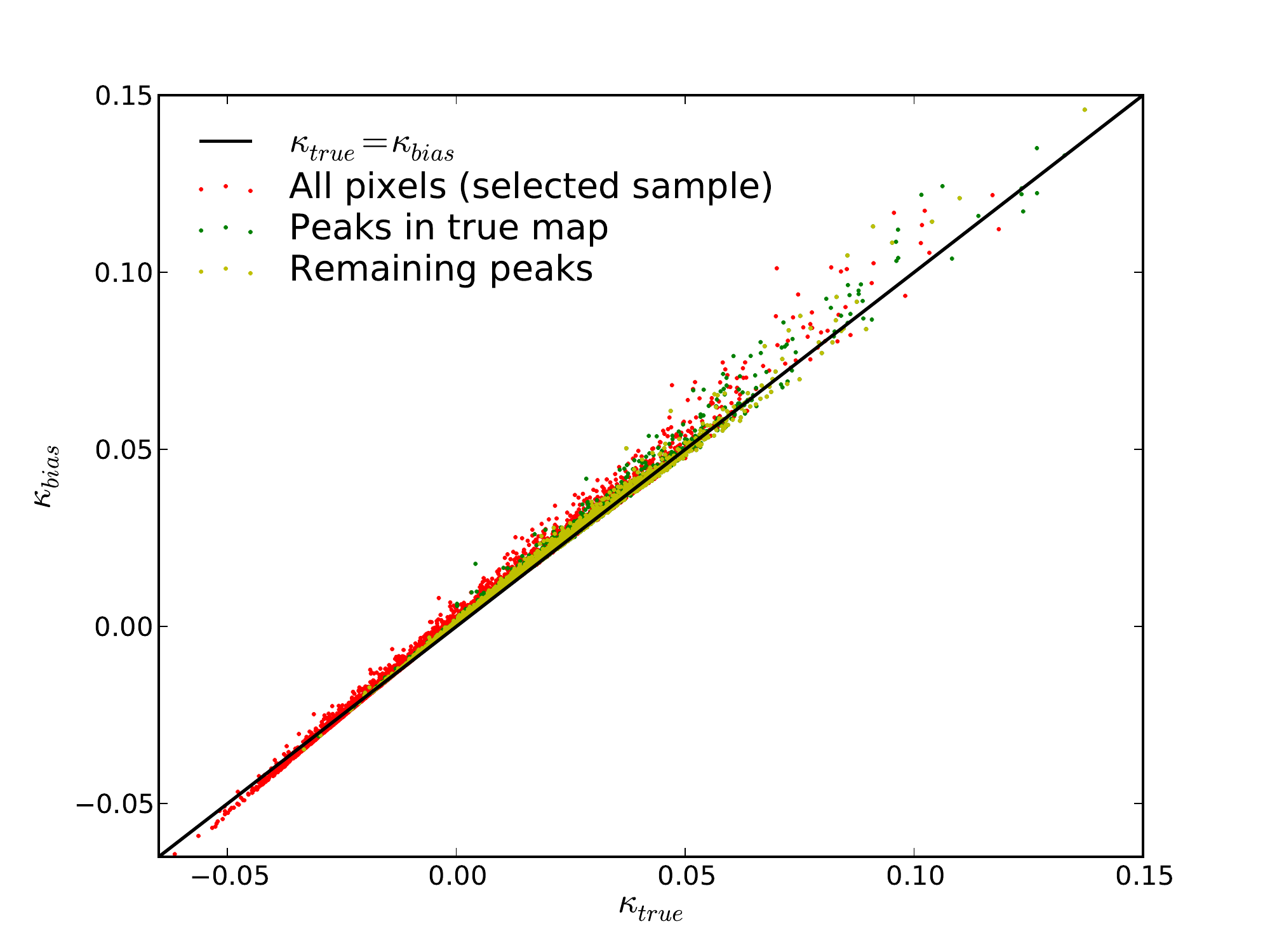}
\caption{\label{kappa scatter} Comparison of $\kappa_{true}$ and
  $\kappa_{bias}$ on a pixel-by-pixel basis in the $s=1.5$ case. A
  random subset (10,000 pixels) of all $2048\times2048$ pixels (red
  dots), pixels that are peaks in ``true'' maps (green dots) and the
  pixels that remain peaks in the ``bias'' maps (yellow dots) are
  shown. Most positive pixels are boosted to a
  higher value. The true peaks that remain peaks in the biased map
  tend to have smaller increases in $\kappa$ than a random pixel. This
  can be attributed to the fact that most such ``survivor'' peaks are
  more dominant -- i.e. stand out more in their local environment
  within a smoothing scale. }
\end{figure}

We have found that MB results in a monotonic increase or decrease for
all $\kappa$ peaks {\em before smoothing}, depending on the sign of
$5s-2$. Fig.~\ref{kappa scatter} shows the change in $\kappa$ values
for all individual pixels, as well as for the peaks, for the $s=1.5$
case. The peaks that survived the MB (the pixels that are peaks in
both ``true'' and ``bias'' maps) tend to have a smaller increase in
their $\kappa$ value than other random pixels. We speculate that these
are the local dominating peaks that could not gain a higher value due
to the lack of higher peaks around them. Interestingly,
Fig.~\ref{kappa scatter} also show a clear cutoff at $\kappa
\lesssim -0.03$ below which no peaks are seen.\footnote{This
  $\kappa_{min}$ could potentially be a cosmological probe, in analogy
  with the cosmology-dependent minimum in the probability distribution
  of $\kappa$ in random directions on the sky \cite{Linder2008}.}

Fig.~\ref{peakcounts} shows that the changes caused by variations
in cosmological parameters tend to be more symmetric in the two
wings of the peak distribution. For example, at $w=-0.8$, we see fewer
high--$\kappa$ peaks, as well as fewer low--$\kappa$ peaks. This shows
that no single cosmological parameter can mimic the changes caused by
MB -- however, a linear combination of the three parameters may still
resemble such change and can be degenerate with the effects of MB (as
we will see below).

Examining the changes due to $\Omega_m$ and $\sigma_8$, we see a clear
degeneracy between these two parameters. This previously known issue
(e.g. \cite{Rozo2010,Basilakos2010,Kratochvil2010,Dietrich2010}) is
similar to that from cluster counts -- both $\Omega_m$ and $\sigma_8$
can change the number of massive halos; therefore, we can obtain the
same number of massive halos (hence the same peak distribution) for a
higher value of $\sigma_8$, as long as we decrease $\Omega_m$. A
product of the two parameters in the form of $\Omega_m\sigma_8^\gamma$
is much more tightly constrained by a fixed number of halos. The value
of $\gamma$ depends on the relevant mass scale being measured, and
varies from 0.3 to 0.6
\cite{Rozo2010,Mantz2008,Henry2009,Vikhlinin2009,Kilbinger2013}. From
our error ellipse, we found $\gamma=0.62$ for the power spectrum and
$\gamma=0.48$ for peak counts, by minimizing
$\Delta\sigma_8/\sigma_8+\gamma\Delta\Omega_m/\Omega_m$ for the 1,000
fitted fiducial maps ($\Delta\sigma_8$ and $\Delta\Omega_m$ are the
differences between the fitted values for an individual map and the
fiducial parameters).

\subsection{Cosmological Parameters}

We are now ready to show that without taking into account the effect
of magnification bias, WL surveys can deliver cosmological parameters
that are biased from the true values by many times their statistical error
$\sigma$ -- for both the power spectrum and peak counts.

\begin{figure} 
\includegraphics[scale=0.45]{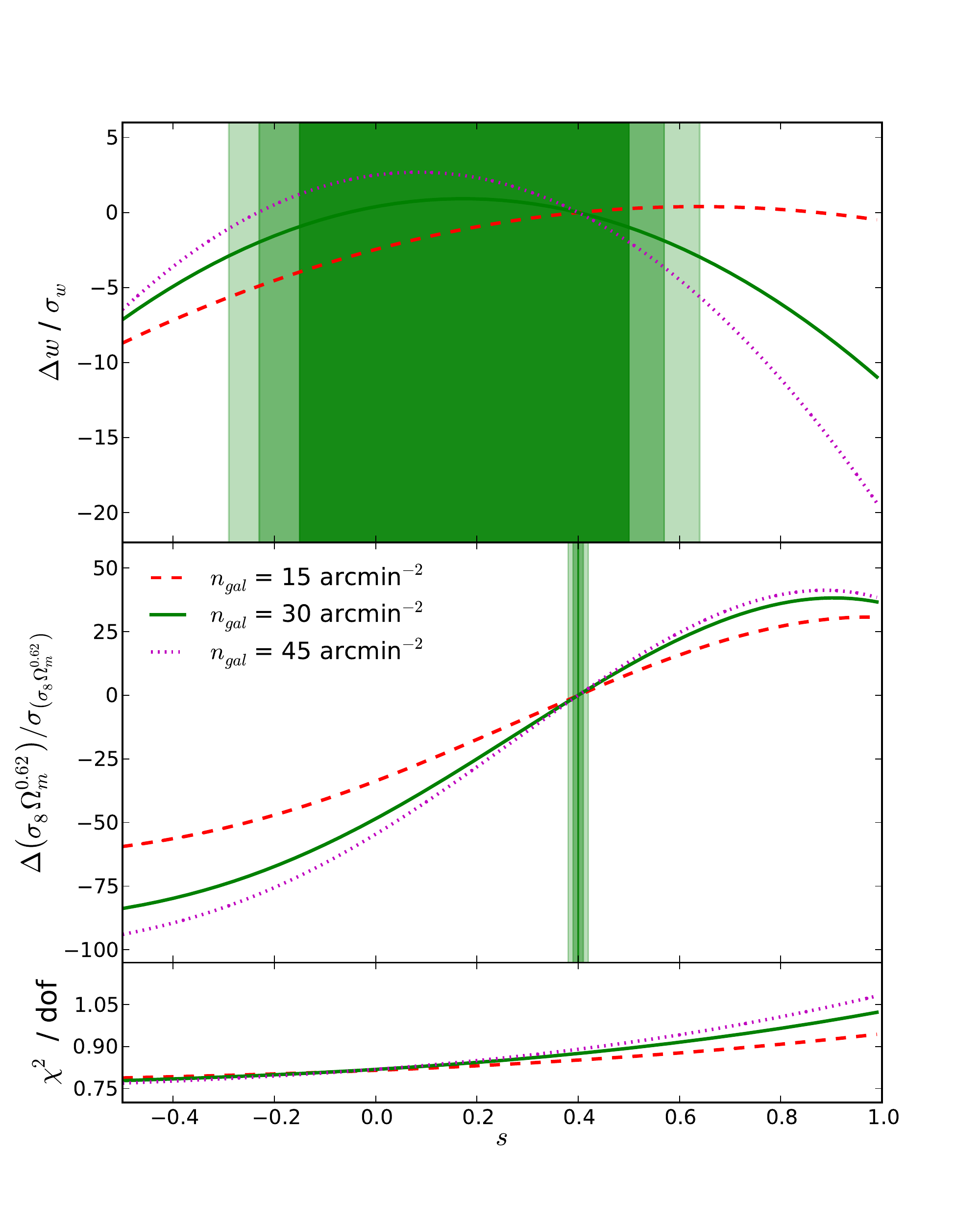}
\caption{\label{del_w_powspec} The biases in cosmological parameters
  inferred from the power spectrum, in units of their standard
  deviation ($\sigma_w=0.016,\sigma_{(\sigma_8\Omega_m^{0.62})}=0.0007$). 
  The shaded regions indicate values of $s$ where the
  cosmology bias is within 1, 2, and 3$\sigma$ (dark to light) for
  $n_{gal} = 30$ arcmin$^{-2}$ (LSST's expected galaxy surface
  density). In the case of pure MB ($s=0.2$) and MB + SB ($s=0.8$) for
  LSST, $w$ is biased by $0.9\sigma$ and $-6.1\sigma$, respectively.
  The best-constrained combination of $\sigma_8\Omega_m^{0.62}$ is
  biased by more than $20\sigma$ in both cases. The error-bar $\sigma$
  has been scaled from our simulation (12 deg$^2$) to LSST's planned
  sky coverage of 20,000 deg$^2$.}
\end{figure}

\begin{figure} 
\includegraphics[scale=0.45]{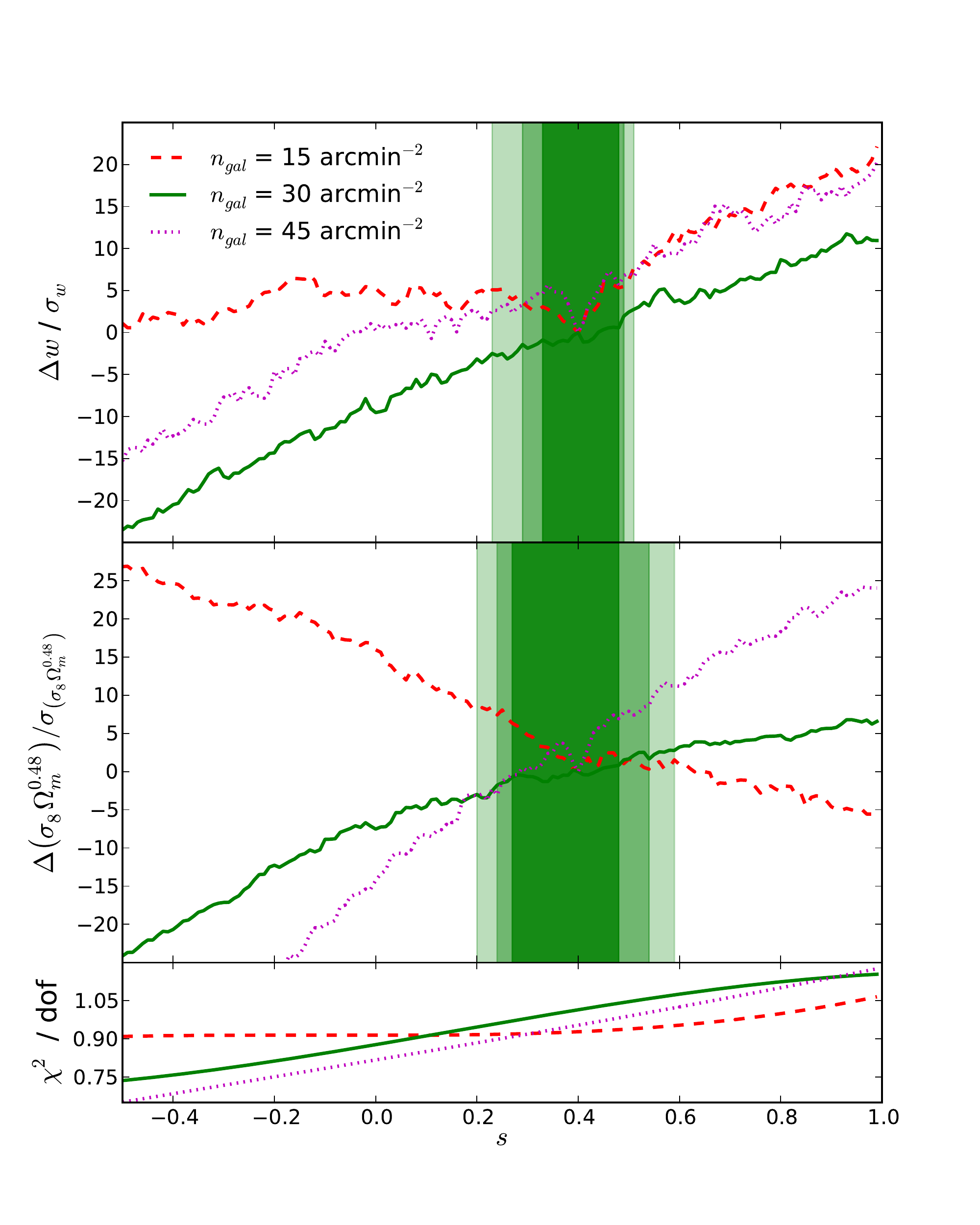}
\caption{\label{del_w} The biases in cosmological parameters inferred
  from the peak counts, in units of their
  standard deviation ($\sigma_w=0.006,\sigma_{(\sigma_8\Omega_m^{0.48})}=0.0004$).
  The shaded regions indicate values of $s$ where
  the cosmology bias is within 1, 2, and 3$\sigma$ (dark to light)
  for $n_{gal} = 30$ arcmin$^{-2}$ (LSST's expected galaxy surface density).
  $w$ is biased by $-3.1\sigma$ ($s=0.2$) and $8.7\sigma$ ($s=0.8$),
  and the combination $\sigma_8\Omega_m^{0.48}$ by $-3.0\sigma$  ($s=0.2$)
  and $4.7\sigma$ ($s=0.8$).  The error-bar $\sigma$ has
  been scaled from our simulation (12 deg$^2$) to LSST's planned sky
  coverage of 20,000 deg$^2$.}
\end{figure}

\begin{figure} 

  \includegraphics[scale=0.45]{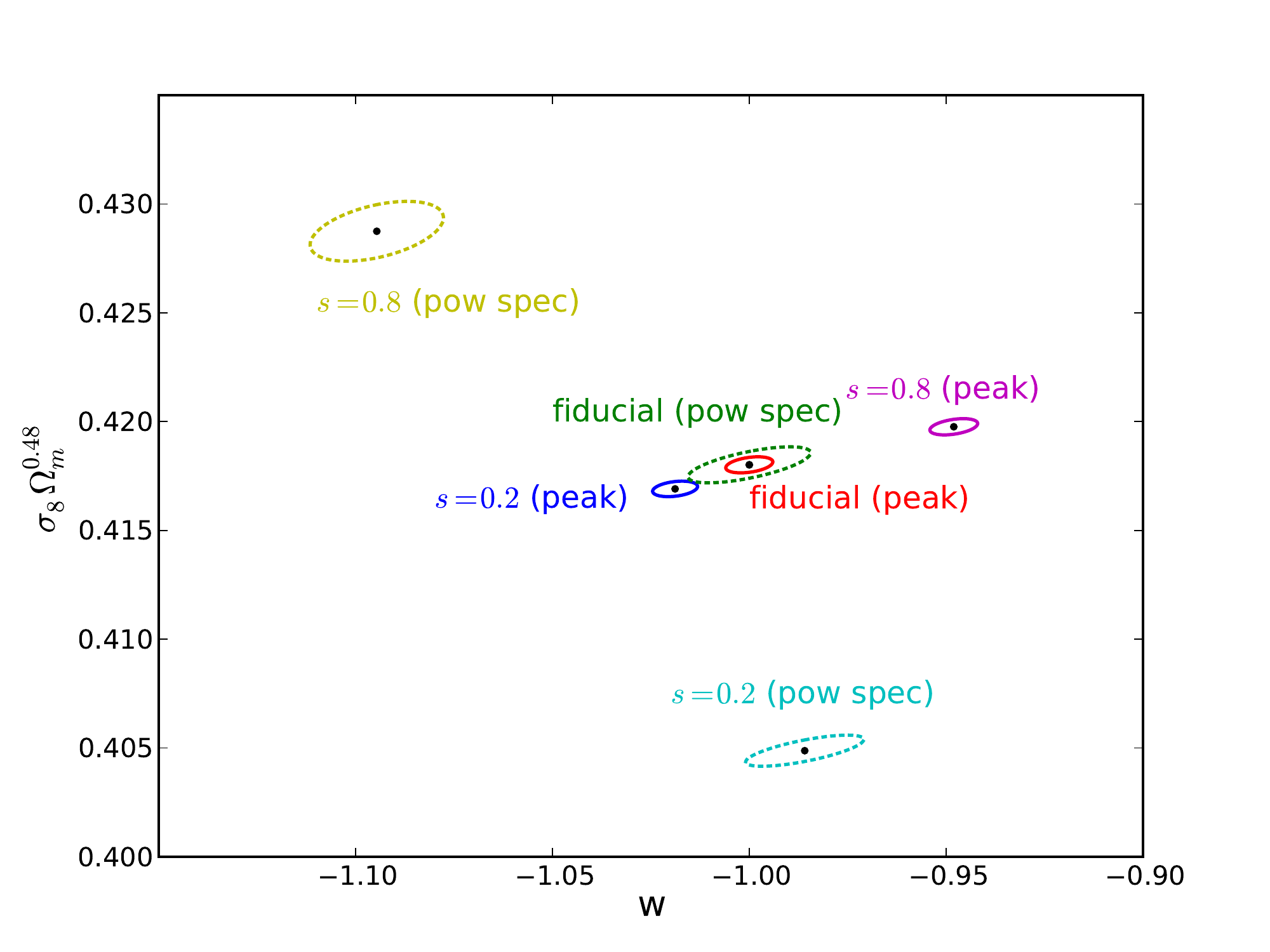}
  \caption{\label{err_ells} Error ellipses for the fiducial (unbiased)
    maps and in the case of magnification bias with $s=0.2$ and $0.8$, for both
    the power spectrum (using $100<\ell<20,000$) and the peak counts
    (with 200 convergence bins and smoothing scale $1/\sqrt 2$ arcmin). Error ellipses contain
    68\% of the best-fits, and have been scaled to LSST's sky coverage
    of 20,000 deg$^2$.}
\end{figure}

Fig.~\ref{del_w_powspec} shows the average deviation of fitted
parameters using the power spectrum, in units of their standard
deviation ($\sigma_w=0.016,\sigma_{(\sigma_8\Omega_m^{0.62})}=0.0007$).
We have computed this cosmology bias for the range of
$-0.5\leq s\leq 1.0$.  The standard deviation is calculated over the
1,000 fiducial maps. Each fitted parameter is marginalized over the
other parameters, and scaled from our simulation (12 deg$^2$) to
LSST's planned sky coverage of 20,000 deg$^2$. The shaded region
indicates the values of $s$ where the deviation of the derived
parameter is within 1$\sigma$, 2$\sigma$ and 3$\sigma$ (dark to
light), for galaxy density $n_{gal}=30$~arcmin$^{-2}$. For $s=0.2$,
$\Delta w/\sigma_w=0.9$ (although interestingly, as shown in the
figure, the bias is not monotonic in $s$) and
${\Delta}(\sigma_8\Omega_m^{0.62})/\sigma_{(\sigma_8\Omega_m^{0.62})}=-25.0$
at 1$\sigma$. We choose to plot $\sigma_8\Omega_m^{0.62}$ instead of
$\sigma_8$ and $\Omega_m$ individually, because the former is much
more tightly constrained, as discussed in \S~\ref{Peakcounts}.  

In Fig.~\ref{del_w}, we show the deviations of cosmological
parameters inferred from the peak counts 
($\sigma_w=0.006,\sigma_{(\sigma_8\Omega_m^{0.48})}=0.0004$). For $s=0.2$, we find $\Delta
w/\sigma_w=-3.1$, $\sim3$ times larger in magnitude than from the power
spectrum; ${\Delta}(\sigma_8\Omega_m^{0.48}) /
\sigma_{(\sigma_8\Omega_m^{0.48})}=-3.0$ at 1$\sigma$, which, on the
other hand, is much lower than from the power spectrum.
For $s=0.8$, we see deviations at similar magnitude but in opposite
directions to the $s=0.2$ case, for both the power spectrum and peak counts.

The biases are again shown in two dimensions in Fig.~\ref{err_ells}, where
the Monte Carlo error ellipses, enclosing 68\% of the best-fits, are explicitly
shown  for the fiducial unbiased maps and biased maps ($s=0.2,0.8$).
In conclusion, WL observations in a survey as large as LSST will need
to take MB into account, by including it in the modeling when fitting
the observations. Combining information from both the power spectrum
and the peak counts will be useful, as these two observables are
impacted by MB in different ways, and their combination can help
mitigate the biases. The value of $s$ (or other parameters describing
higher-order lensing corrections) could be potentially additional
parameters in a fitting procedure, simultaneously with the
cosmological parameters. We expect that MB has a smaller impact on the current surveys, mainly
due to their smaller sky coverage (e.g. COSMOS: $\approx$2 deg$^2$,
CFHTLenS: 150 deg$^2$). After scaling $\sigma$ by their sky coverage,
we found the deviations to be of order $\sim 0.01\sigma$ for COSMOS,
and $\sim 0.1\sigma$ for CFHTLenS.  

The observed galaxy number density will also affect the level of
MB. In Figures~\ref{del_w_powspec} and~\ref{del_w}, we also show the
parameter biases for $n_{gal}=15$ and 45 arcmin$^{-2}$. For the power spectrum, 
the slope near $s=0.4$ tends to be steeper for larger $n_{gal}$. This means deeper
surveys with higher galaxy number densities (hence smaller galaxy
noise) are more sensitive to MB when galaxy noise (eq.~\ref{eq:galaxynoise}) is
smaller. For peak counts, MB impacts the derived $\sigma_8\Omega_m^{0.48}$ for
 shallower surveys ($n_{gal}=15$ arcmin) in opposite
direction to surveys with higher number density.

\section{Discussions}\label{Discussions}

In this work, we made several assumptions and simplifications, which
we must highlight here:

(i) We assumed the number of peaks and the power spectrum depends
linearly on cosmology. For example, in our analysis, we used
``forward'' derivatives for $d\bar{N}/dp$, built with the fiducial and the
three ``high'' cosmologies from Table~\ref{tab:Cosmologies} for a
finite difference. We can also use the three ``low'' cosmologies to
obtain ``backward'' finite-difference derivatives. When we do so, 
we find the resulting deviations to have similar magnitude (Table~\ref{tab:fwbw})
to the ones from ``forward'' derivatives, 
except for a significantly lower value for $\Delta\sigma_8\Omega_m^{0.48}$
for peak counts.

\begin{table}
\begin{tabular}{ccccc} 
\hline					
	&	\multicolumn{2}{c}{	$s = 0.2$	}	&	\multicolumn{2}{c}{	$s = 0.8$	}	\\
\hline									
power spectrum	&	forward	&	backward	&	forward	&	backward	\\
\cline{2-5}									
$\Delta \Omega_m/\sigma_{\Omega_m}$	&	-21.0	&	-23.2	&	61.6	&	67.0	\\
$\Delta w/\sigma_w$	&	0.9	&	0.4	&	-6.1	&	-4.1	\\
$\Delta \sigma_8/\sigma_{\sigma_8}$	&	7.6	&	8.6	&	-28.3	&	-30.9	\\
${\Delta}(\sigma_8\Omega_m^{0.62}) / \sigma_{(\sigma_8\Omega_m^{0.62})}$	&	-25.0	&	-27.3	&	36.0	&	37.4	\\
\hline									
peak counts	&	forward	&	backward	&	forward	&	backward	\\
\cline{2-5}									
$\Delta \Omega_m/\sigma_{\Omega_m}$	&	1.4	&	1.5	&	-2.9	&	-3.2	\\
$\Delta w/\sigma_w$	&	-3.1	&	-4.4	&	8.7	&	4.5	\\
$\Delta \sigma_8/\sigma_{\sigma_8}$	&	-2.6	&	-2.7	&	4.8	&	3.1	\\
${\Delta}(\sigma_8\Omega_m^{0.48}) / \sigma_{(\sigma_8\Omega_m^{0.48})}$	&	-3.0	&	-3.1	&	4.7	&	0.2	\\
\hline									
\end{tabular}
\caption[]{\label{tab:fwbw} Deviations of cosmological parameters
 evaluated at $s=0.2$ (MB only) and $s=0.8$ (MB + SB). Results 
from ``forward'' and ``backward'' derivatives are compared side-by-side.}
\end{table}

We also attempted to use a spline interpolation, using all three
data-points for each parameter to describe the cosmology-dependence.
This enables us to utilize all 7 cosmologies simultaneously, but we
lose the advantage of the analytical method to obtain the best-fits
(eq.~\ref{analytical dp}). We used the numerical method to find the
best-fits with spline interpolation, and found mean biases similar to
those from linear interpolations. However, the error ellipses from
spline interpolation were considerably smaller, and suspiciously
coincident with our simulated range of model parameters. This is
likely due to the spline tails that curve dramatically outside our
parameter region, and hence artificially force the fit to stay within
our simulated range of model parameters for each map. To solve this
issue, we will need to have a larger grid of simulation parameters,
which will also help us understand the dependence of peak counts on
cosmology more accurately.

(ii) We used convergence maps only at a single redshift. This is
motivated by the fact that the effect of MB depends weakly on $z$. 
At low redshift, $s$
is mainly dependent on the slope of this power-law tail, and MB will
have similar level of impact for all galaxies. For galaxies at higher
redshift, $m_{\rm lim}$, when redshifted to the rest frame of the
galaxy, moves closer to the exponential part of the luminosity
function, so we expect $s$ to increase to a larger value. A
redshift-dependent correction to MB that folds in the correct
$z$-distribution of high-$z$ galaxies will eventually be necessary.

(iii) We ignored all instrumental and measurement errors. In reality,
the point spread function (PSF) deconvolution and the measurement of
galaxy shapes accurately is a difficult task, and has received
thorough discussions
(e.g.\cite{Kaiser1995,Bonnet1995}). Ref.~\cite{Bard2013} used
simulated shear maps with realistic galaxy properties and has taken
into account distortions from both the atmosphere and optical errors
expected for LSST. They have shown that, though peak significance is
reduced, the addition of these errors does not significantly degrade
the cosmological constraints, compared to considering shape noise
only. While our basic conclusion, that MB is significant, likely
remains valid in the presence of such errors, the detailed modeling of
MB will need to incorporate these additional sources of error.

(iv) In this paper, we choose to work with convergence maps, as they
are computationally simpler.
Using galaxies with sizes larger than the PSF, the convergence field
can potentially be inferred by combining galaxy size and flux measurements, as
lensing modifies these two quantities by different factors of
$1+\kappa$ and $1+2\kappa$, respectively, in the weak lensing limit
\cite{Vallinotto2011, Schmidt2012, Casaponsa2013, Heavens2013}.  In
current practice, reduced shear maps are obtained by measuring the shapes of
individual galaxies. The observer can deduce the aperture mass
($M_{ap}$), a smoothed form of convergence, by applying a convolution
over tangential components of
shear \cite{Schneider1996}. Ref.~\cite{Pires2012} has shown that both shear
and convergence statistics give similar constraints when compared at
the same scale, but once again, the impact of lensing bias should be
modeled directly on the shear field.

(v) Although WL surveys may implement a sharp flux cut, the size bias
is likely to be more complicated, with an effective weighting on
galaxies that depends monotonically on their size, but in a gradual
fashion, rather than a step function.  In the idealized case of a
sharp size cut, our analysis remains applicable, with a suitable
re-interpretation of $5s$ as a stand-in for $5s+\beta$, where $\beta$
is the logarithmic slope of the size distribution (at the size
cut). In this simplified case, the bias induced by the size cut is
likely larger than the one induced by the flux cut. For example,
\cite{Schmidt2009b} showed that, for a survey with magnitude cut
$i_{AB}=24$ and size cut $r=1.2"$, the impact of MB becomes positive
and $q=5s+\beta-2\sim1-2$, which is equivalent to our cases with
$s=0.6-0.8$. From Fig.~\ref{err_ells}, we see the derived parameters 
remain many $\sigma$ away from the true parameters, but in the opposite
direction. This demonstrates that size cut is likely to be important,
and also that it is necessary in future work to investigate the
effects of size bias in more detail.

(vi) We have tested the impact of MB on three parameters, $\Omega_m$,
$w$ and $\sigma_8$. When additional cosmological parameters
are considered (e.g. $\Omega_b$, $H_0$, $n_s$, $w_a$), the impact of
MB may be more severe, since a combination involving the new
parameters could mimic the MB better. In order to test this, we need
to run more N-body simulations with other parameters varying to build
a more complete cosmological model.

\begin{figure} 
  \includegraphics[scale=0.45]{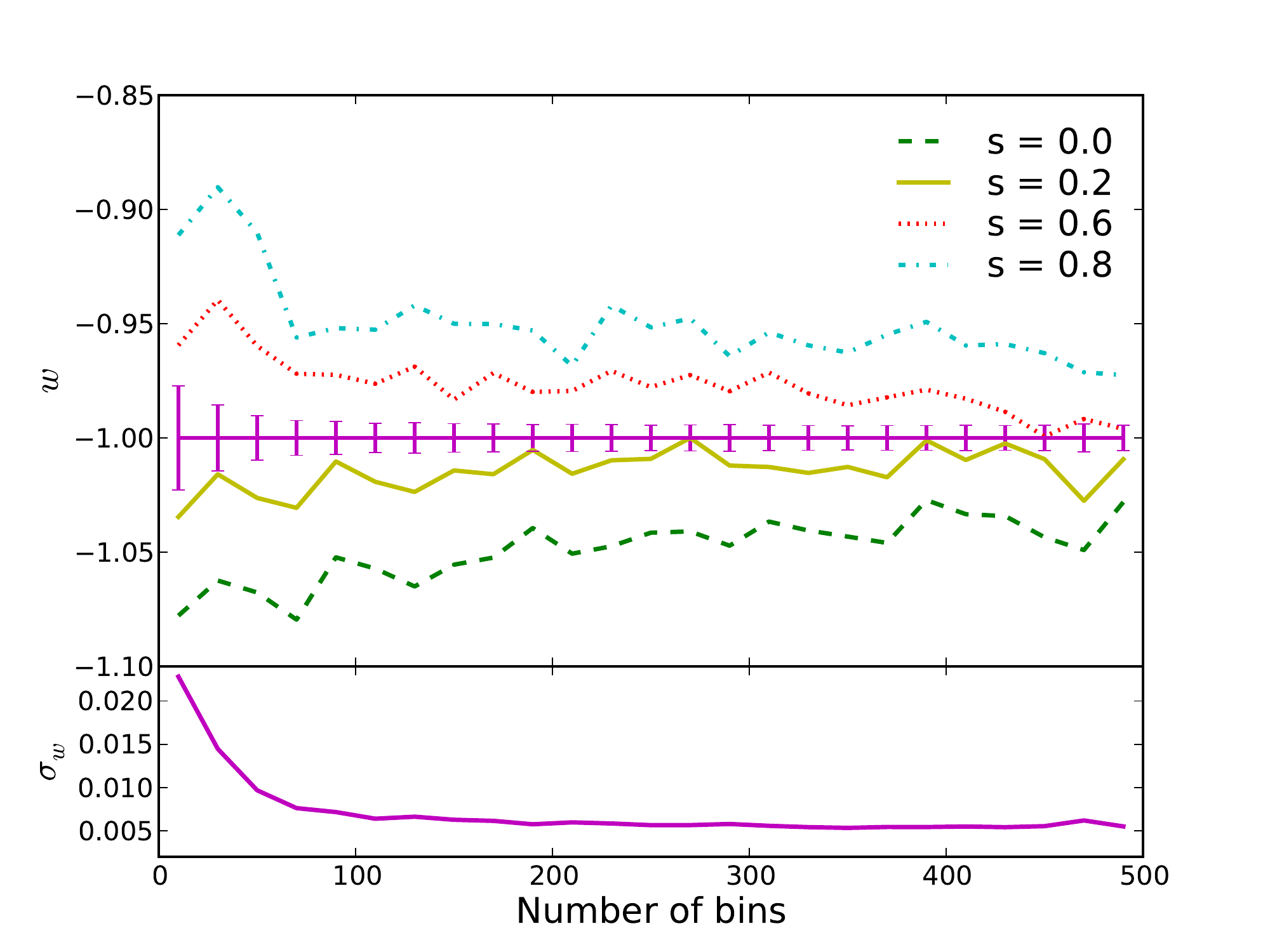}
  \includegraphics[scale=0.45]{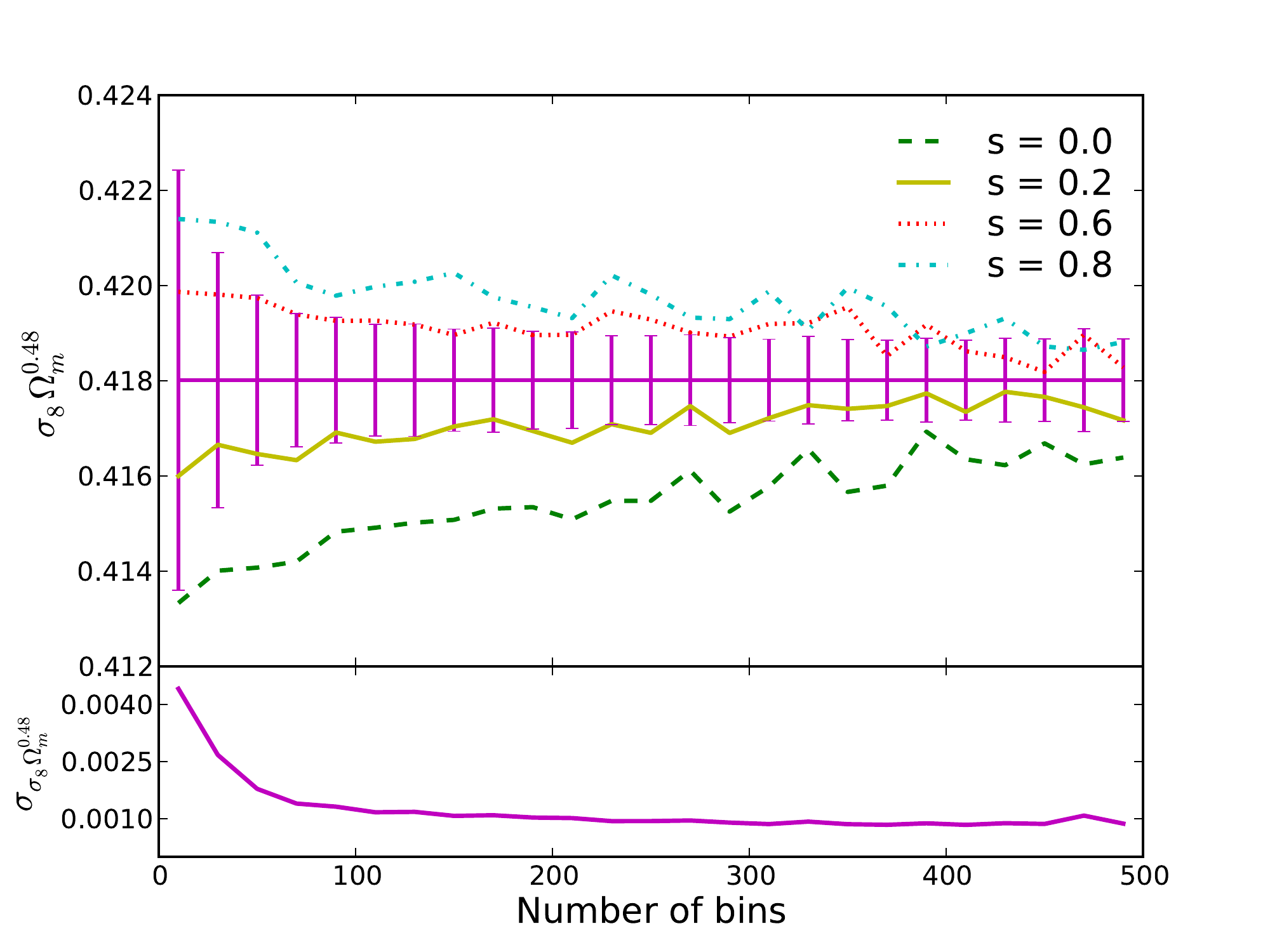}
  \caption{\label{err_bins} The derived cosmological parameters and
    marginalized error for each parameter as a function of the number
    of bins, using peak counts. The error sizes have been scaled to
    LSST's sky coverage of 20,000 deg$^2$. The error sizes tend to
    decrease with larger number of bins. The results for $\gsim500$
    bins are unreliable due to the limited number of realizations in
    our simulation suite.}
\end{figure}

(vii) Optimizing the number of bins has not been the focus of this
work.  However, the choice of the number of bins has an effect on the
error sizes. As shown in Fig.~\ref{err_bins}, for peak counts, the
values of derived parameters and marginalized errors only converge at
$\gsim150$ bins. Once the number of bins exceeds this value, we see a
roughly constant plateau extending to $\gsim500$ bins (beyond which
the results become unreliable, due to having too few realizations of
maps and the sample covariance matrix becoming singular near 1000 bins).
Therefore, we chose to use 200 bins in this work.

\section{Summary}\label{Summary}

In this paper, we have studied the effect of magnification bias on
peak statistics, using convergence maps from ray-tracing N-body
simulations. Using maps in a suite of simulations, we can predict the
convergence power spectrum or peak count distribution as a function of
$\Omega_m, w$, and $\sigma_8$. Using this tool, we found the biases in
cosmological parameters, when convergence maps in the fiducial
cosmology, modified by magnification bias, were used to find the
best-fit cosmology, without taking MB into account in the fits.

Near the flux limit of future WL survey, such as LSST, the galaxy number counts
have a logarithmic slope of $s\approx0.2$. This causes a bias in the
inferred value of $w$ by $0.9\sigma$ and of $\sigma_8\Omega_m^{0.62}$
by $-25.0\sigma$ when using the power spectrum, and by $-3.1\sigma$ for
$w$ and $-3.1\sigma$ for $\sigma_8\Omega_m^{0.48}$ when using peak
counts. These results are scaled to WL observations expected from
LSST. However, for recent surveys, such as COSMOS and CFHTLenS, the
deviations are generally negligible ($\sim 0.01\sigma$ and $\sim
0.1\sigma$, respectively), due to their smaller sky coverage. 

We conclude that it is necessary that cosmological simulations
consider MB effects, when they are used to match observations. We have
found that $w$ inferred from the power spectrum is less impacted by
MB, but peak count is a less biased method to infer
$\sigma_8\Omega_m^\gamma$. Future work on magnification biases should
incorporate the many improvements we have emphasized that are necessary,
including (i) the redshift-dependence of the bias, (ii) the impact on
shear maps with realistic measurement errors and the peak statistics
derived from these maps, (iii) more complex biases induced by the
size-dependent measurement errors cut on galaxies, and (iv)
additionally, the potential of using magnification bias and size bias
as a signal to tighten the constraints on convergence field
\cite{Vallinotto2011, Schmidt2012, Casaponsa2013, Heavens2013}.  Our
results suggest that lensing biases can be mitigated by combining the
power spectrum and the peak counts, which produce biases in very
different directions in cosmological parameter space.
 
\begin{acknowledgments}
  We thank Kevin Huffenberger for useful discussions.
  This research utilized resources at the New York Center for
  Computational Sciences, a cooperative effort between Brookhaven
  National Laboratory and Stony Brook University, supported in part by
  the State of New York. This work is supported in part by the
  U.S. Department of Energy under Contract No. DE-AC02-98CH10886 and
  by the NSF under grant AST-1210877.  The simulations were created on
  the IBM Blue Gene/L and /P New York Blue computer and the maps were
  created analyzed on the LSST/Astro Linux cluster at BNL.

\end{acknowledgments}

\bibliographystyle{physrev}

\begin{thebibliography}{10}

\bibitem{Refregier2003}
A.~{Refregier},
\newblock \araa {\bf 41}, 645 (2003), [arXiv:arXiv:astro-ph/0307212].

\bibitem{Schneider2005}
P.~{Schneider},
\newblock ArXiv Astrophysics e-prints  (2005), [arXiv:arXiv:astro-ph/0509252].

\bibitem{Hoekstra2008}
H.~{Hoekstra} and B.~{Jain},
\newblock Annual Review of Nuclear and Particle Science {\bf 58}, 99 (2008),
  [arXiv:0805.0139].

\bibitem{Bartelmann2010}
M.~{Bartelmann},
\newblock Classical and Quantum Gravity {\bf 27}, 233001 (2010).

\bibitem{Schrabback2010}
T.~{Schrabback} {\em et~al.},
\newblock \aap {\bf 516}, A63 (2010), [arXiv:0911.0053].

\bibitem{Kilbinger2013}
M.~{Kilbinger} {\em et~al.},
\newblock \mnras {\bf 430}, 2200 (2013), [arXiv:1212.3338].

\bibitem{Webster1988}
R.~L. {Webster}, P.~C. {Hewett}, M.~E. {Harding} and G.~A. {Wegner},
\newblock \nat {\bf 336}, 358 (1988).

\bibitem{Fugmann1988}
W.~{Fugmann},
\newblock \aap {\bf 204}, 73 (1988).

\bibitem{Narayan1989}
R.~{Narayan},
\newblock \apjl {\bf 339}, L53 (1989).

\bibitem{Schneider1989}
P.~{Schneider},
\newblock \aap {\bf 221}, 221 (1989).

\bibitem{Villumsen1995}
J.~{Verner Villumsen},
\newblock ArXiv Astrophysics e-prints  (1995), [arXiv:arXiv:astro-ph/9512001].

\bibitem{Villumsen1997}
J.~V. {Villumsen}, W.~{Freudling} and L.~N. {da Costa},
\newblock \apj {\bf 481}, 578 (1997), [arXiv:arXiv:astro-ph/9606084].

\bibitem{Moessner1998}
R.~{Moessner}, B.~{Jain} and J.~V. {Villumsen},
\newblock \mnras {\bf 294}, 291 (1998), [arXiv:arXiv:astro-ph/9708271].

\bibitem{Kaiser1998}
N.~{Kaiser},
\newblock \apj {\bf 498}, 26 (1998), [arXiv:arXiv:astro-ph/9610120].

\bibitem{Loverde2007}
M.~{Loverde}, L.~{Hui} and E.~{Gazta{\~n}aga},
\newblock \prd {\bf 75}, 043519 (2007), [arXiv:arXiv:astro-ph/0611539].

\bibitem{Matsubara2000}
T.~{Matsubara},
\newblock \apjl {\bf 537}, L77 (2000), [arXiv:arXiv:astro-ph/0004392].

\bibitem{Hui2007}
L.~{Hui}, E.~{Gazta{\~n}aga} and M.~{Loverde},
\newblock \prd {\bf 76}, 103502 (2007), [arXiv:0706.1071].

\bibitem{Hui2008}
L.~{Hui}, E.~{Gazta{\~n}aga} and M.~{Loverde},
\newblock \prd {\bf 77}, 063526 (2008), [arXiv:0710.4191].

\bibitem{Loverde2010}
M.~{Loverde}, S.~{Marnerides}, L.~{Hui}, B.~{M{\'e}nard} and A.~{Lidz},
\newblock \prd {\bf 82}, 103507 (2010), [arXiv:1004.1165].

\bibitem{Schmidt2009}
F.~{Schmidt}, E.~{Rozo}, S.~{Dodelson}, L.~{Hui} and E.~{Sheldon},
\newblock \apj {\bf 702}, 593 (2009), [arXiv:0904.4703].

\bibitem{Schmidt2009b}
F.~{Schmidt}, E.~{Rozo}, S.~{Dodelson}, L.~{Hui} and E.~{Sheldon},
\newblock Physical Review Letters {\bf 103}, 051301 (2009), [arXiv:0904.4702].

\bibitem{Schmidt2011}
F.~{Schmidt} and E.~{Rozo},
\newblock \apj {\bf 735}, 119 (2011), [arXiv:1009.0757].

\bibitem{DETF}
A.~{Albrecht} {\em et~al.},
\newblock arXiv:astro-ph/0609591.

\bibitem{JV00}
B.~{Jain} and L.~{Van Waerbeke},
\newblock \apjl {\bf 530}, L1 (2000), [arXiv:astro-ph/9910459].

\bibitem{Dietrich2010}
J.~P. {Dietrich} and J.~{Hartlap},
\newblock \mnras {\bf 402}, 1049 (2010), [arXiv:0906.3512].

\bibitem{Maturi2010}
M.~{Maturi}, C.~{Angrick}, F.~{Pace} and M.~{Bartelmann},
\newblock \aap {\bf 519}, A23 (2010), [arXiv:0907.1849].

\bibitem{Kratochvil2010}
J.~M. {Kratochvil}, Z.~{Haiman} and M.~{May},
\newblock \prd {\bf 81}, 043519 (2010), [arXiv:0907.0486].

\bibitem{Yang2011}
X.~{Yang} {\em et~al.},
\newblock \prd {\bf 84}, 043529 (2011), [arXiv:1109.6333].

\bibitem{Marian2012}
L.~{Marian}, R.~E. {Smith}, S.~{Hilbert} and P.~{Schneider},
\newblock \mnras {\bf 423}, 1711 (2012), [arXiv:1110.4635].

\bibitem{Kratochvil2012}
J.~M. {Kratochvil} {\em et~al.},
\newblock \prd {\bf 85}, 103513 (2012), [arXiv:1109.6334].

\bibitem{Pires2012}
S.~{Pires}, A.~{Leonard} and J.-L. {Starck},
\newblock \mnras {\bf 423}, 983 (2012), [arXiv:1203.2877].

\bibitem{Yang2013}
X.~{Yang}, J.~M. {Kratochvil}, K.~{Huffenberger}, Z.~{Haiman} and M.~{May},
\newblock \prd {\bf 87}, 023511 (2013), [arXiv:1210.0608].

\bibitem{Bard2013}
D.~{Bard} {\em et~al.},
\newblock \apj {\bf 774}, 49 (2013), [arXiv:1301.0830].

\bibitem{Turner1984}
E.~L. {Turner}, J.~P. {Ostriker} and J.~R. {Gott}, III,
\newblock \apj {\bf 284}, 1 (1984).

\bibitem{Schechter1976}
P.~{Schechter},
\newblock \apj {\bf 203}, 297 (1976).

\bibitem{Gabasch2004}
A.~{Gabasch} {\em et~al.},
\newblock \aap {\bf 421}, 41 (2004), [arXiv:arXiv:astro-ph/0403535].

\bibitem{Gabasch2006}
A.~{Gabasch} {\em et~al.},
\newblock \aap {\bf 448}, 101 (2006), [arXiv:arXiv:astro-ph/0510339].

\bibitem{Chang2013}
C.~{Chang} {\em et~al.},
\newblock ArXiv e-prints  (2013), [arXiv:1305.0793].

\bibitem{LSSTSciBook2009}
{LSST Science Collaboration} {\em et~al.},
\newblock ArXiv e-prints  (2009), [arXiv:0912.0201].

\bibitem{Euclid2009}
R.~{Laureijs},
\newblock ArXiv e-prints  (2009), [arXiv:0912.0914].

\bibitem{COSMOS2007}
N.~{Scoville} {\em et~al.},
\newblock \apjs {\bf 172}, 38 (2007), [arXiv:arXiv:astro-ph/0612306].

\bibitem{CFHTLS2012}
C.~{Heymans} {\em et~al.},
\newblock \mnras {\bf 427}, 146 (2012), [arXiv:1210.0032].

\bibitem{DES2005}
{The Dark Energy Survey Collaboration},
\newblock ArXiv Astrophysics e-prints  (2005), [arXiv:arXiv:astro-ph/0510346].

\bibitem{DUNE2009}
A.~{Refregier},
\newblock Experimental Astronomy {\bf 23}, 17 (2009), [arXiv:0802.2522].

\bibitem{KiDS2013}
J.~T.~A. {de Jong}, G.~A. {Verdoes Kleijn}, K.~H. {Kuijken} and E.~A.
  {Valentijn},
\newblock Experimental Astronomy {\bf 35}, 25 (2013), [arXiv:1206.1254].

\bibitem{HSC2010}
M.~{Takada},
\newblock {Subaru Hyper Suprime-Cam Project},
\newblock in {\em American Institute of Physics Conference Series}, edited by
  N.~{Kawai} and S.~{Nagataki}, , American Institute of Physics Conference
  Series Vol. 1279, pp. 120--127, 2010.

\bibitem{Komatsu2011}
E.~{Komatsu} {\em et~al.},
\newblock \apjs {\bf 192}, 18 (2011), [arXiv:1001.4538].

\bibitem{Lewis2000}
A.~{Lewis}, A.~{Challinor} and A.~{Lasenby},
\newblock \apj {\bf 538}, 473 (2000), [arXiv:arXiv:astro-ph/9911177].

\bibitem{Hockney1988}
R.~Hockney and J.~Eastwood,
\newblock Chap {\bf 4}, 113 (1988).

\bibitem{Song2004}
Y.-S. {Song} and L.~{Knox},
\newblock \prd {\bf 70}, 063510 (2004), [arXiv:arXiv:astro-ph/0312175].

\bibitem{vanWaerbeke2000}
L.~{van Waerbeke},
\newblock \mnras {\bf 313}, 524 (2000), [arXiv:arXiv:astro-ph/9909160].

\bibitem{Limber1953}
D.~N. {Limber},
\newblock \apj {\bf 117}, 134 (1953).

\bibitem{Smith2003}
R.~E. {Smith} {\em et~al.},
\newblock \mnras {\bf 341}, 1311 (2003), [arXiv:arXiv:astro-ph/0207664].

\bibitem{Hartlap2007}
J.~{Hartlap}, P.~{Simon} and P.~{Schneider},
\newblock \aap {\bf 464}, 399 (2007), [arXiv:arXiv:astro-ph/0608064].

\bibitem{Anderson2003}
T.~W. {Anderson},
\newblock {\em {An Introduction to Multivariate Statistical Analysis}}, third
  ed. (Wiley, New York, NY, 2003).

\bibitem{Linder2008}
E.~V. {Linder},
\newblock \jcap {\bf 3}, 19 (2008), [arXiv:0711.0743].

\bibitem{Rozo2010}
E.~{Rozo} {\em et~al.},
\newblock \apj {\bf 708}, 645 (2010), [arXiv:0902.3702].

\bibitem{Basilakos2010}
S.~{Basilakos} and M.~{Plionis},
\newblock \apjl {\bf 714}, L185 (2010), [arXiv:1003.2559].

\bibitem{Mantz2008}
H.~{Mantz}, K.~{Jacobs} and K.~{Mecke},
\newblock Journal of Statistical Mechanics: Theory and Experiment {\bf 12}, 15
  (2008).

\bibitem{Henry2009}
J.~P. {Henry}, A.~E. {Evrard}, H.~{Hoekstra}, A.~{Babul} and A.~{Mahdavi},
\newblock \apj {\bf 691}, 1307 (2009), [arXiv:0809.3832].

\bibitem{Vikhlinin2009}
A.~{Vikhlinin} {\em et~al.},
\newblock \apj {\bf 692}, 1060 (2009), [arXiv:0812.2720].

\bibitem{Kaiser1995}
N.~{Kaiser}, G.~{Squires} and T.~{Broadhurst},
\newblock \apj {\bf 449}, 460 (1995), [arXiv:arXiv:astro-ph/9411005].

\bibitem{Bonnet1995}
H.~{Bonnet} and Y.~{Mellier},
\newblock \aap {\bf 303}, 331 (1995).

\bibitem{Vallinotto2011}
A.~{Vallinotto}, S.~{Dodelson} and P.~{Zhang},
\newblock \prd {\bf 84}, 103004 (2011), [arXiv:1009.5590].

\bibitem{Schmidt2012}
F.~{Schmidt} {\em et~al.},
\newblock \apjl {\bf 744}, L22 (2012), [arXiv:1111.3679].

\bibitem{Casaponsa2013}
B.~{Casaponsa} {\em et~al.},
\newblock \mnras {\bf 430}, 2844 (2013), [arXiv:1209.1646].

\bibitem{Heavens2013}
A.~{Heavens}, J.~{Alsing} and A.~H. {Jaffe},
\newblock \mnras {\bf 433}, L6 (2013), [arXiv:1302.1584].

\bibitem{Schneider1996}
P.~{Schneider},
\newblock \mnras {\bf 283}, 837 (1996), [arXiv:arXiv:astro-ph/9601039].

\end{thebibliography}

\end{document}